\documentclass[modern]{aastex61}

\newcommand{\co}[3]{\ensuremath{\rm{^{#1} CO ({#2}-{#3})}}}

\usepackage[noend]{algpseudocode}
\usepackage{enumitem}

\shorttitle{Auto-multithresh}
\shortauthors{Kepley et al.}

\begin{document}

\title{Auto-multithresh: A General Purpose Automasking Algorithm}

\author{Amanda A. Kepley}
\affiliation{National Radio Astronomy Observatory, 520 Edgemont Road, Charlottesville, VA 22903-2475, USA}

\author{Takahiro Tsutsumi}
\affiliation{National Radio Astronomy Observatory, P.O. Box O, 1003 Lopezville Road, Socorro, NM 87801-0387, USA}

\author{Crystal L. Brogan}
\affiliation{National Radio Astronomy Observatory, 520 Edgemont Road, Charlottesville, VA 22903-2475, USA}

\author{Remy Indebetouw}
\affiliation{National Radio Astronomy Observatory, 520 Edgemont Road, Charlottesville, VA 22903-2475, USA}
\affiliation{Department of Astronomy, University of Virginia, PO Box 400325, Charlottesville, VA 22904, USA}

\author{Ilsang Yoon}
\affiliation{National Radio Astronomy Observatory, 520 Edgemont Road, Charlottesville, VA 22903-2475, USA}

\author{Brian Mason}
\affiliation{National Radio Astronomy Observatory, 520 Edgemont Road, Charlottesville, VA 22903-2475, USA}

\author{Jennifer Donovan Meyer}
\affiliation{National Radio Astronomy Observatory, 520 Edgemont Road, Charlottesville, VA 22903-2475, USA}

\correspondingauthor{Amanda A. Kepley}
\email{akepley@nrao.edu}

\begin{abstract}

Producing images from interferometer data requires accurate modeling of the sources in the field of view, which is typically done using the \textsc{clean} algorithm. Given the large number of degrees of freedom in interferometeric images, one constrains the possible model solutions for \textsc{clean} by masking regions that contain emission. Traditionally this process has largely been done by hand.  This approach is not possible with today's large data volumes which require automated imaging pipelines. This paper describes an automated masking algorithm that operates within \textsc{clean} called \textsc{auto-multithresh}. This algorithm was developed and validated using a set of $\sim1000$ ALMA images chosen to span a range of intrinsic morphology and data characteristics. It takes a top-down approach to producing masks: it uses the residual images to identify significant peaks and then expands the mask to include emission associated with these peaks down to lower signal-to-noise noise.  The \textsc{auto-multithresh} algorithm has been implemented in CASA and has been used in production as part of the ALMA Imaging Pipeline starting with Cycle 5. It has been shown to be able to mask a wide range of emission ranging from simple point sources to complex extended emission with minimal tuning of the parameters based on the point spread function of the data. Although the algorithm was developed for ALMA, it is general enough to have been used successfully with data from other interferometers with appropriate parameter tuning.  Integrating the algorithm more deeply within the minor cycle could lead to future performance improvements.

\end{abstract}

\keywords{radio continuum: general --  radio lines: general -- techniques: image processing --  techniques: interferometric --  submillimeter: general}


\section{Introduction} \label{sec:intro}

The fundamental data products produced by interferometers are visibilities, which are the cross-correlations between pairs of antennas. These visibilities can be inverse Fourier transformed to form images using the van Cittert-Zernike theorem \citep[Chapter 3.1]{Thompson2017InterferometryAstronomy}. The  image  produced by this inversion  -- the so-called ``dirty'' image -- contains artifacts resulting from the discrete sampling of the $u$-$v$ plane by the array. These artifacts are commonly removed prior to the analysis of the image using the \textsc{clean} algorithm (\citealp{Hogbom1974ApertureBaselines,Clark1980AnCLEAN,Schwab1984RelaxingInterferometry}; see also \citealt{Rau2018SynergyAnalysis} for a review of other techniques).  In its most basic form, \textsc{clean} iteratively models the observed emission as a series of point sources. It multiplies these point sources by the point spread function (PSF) of the image -- the ``dirty" beam --  and then subtracts them from the data to produce the residuals. The final image is created by multiplying the point source model  by a Gaussian function with a width similar to that of the central peak of the PSF -- the ``\textsc{clean}'' beam -- and the resulting model image is added to the residuals to create the final image. 

The key to producing an image suitable for scientific analysis with \textsc{clean} is to generate a good model of the source. The degrees of freedom in the model can significantly exceed the number of independent visibilities, particularly for complex sources, leading to non-unique solutions. To avoid unrealistic solutions, the user typically constrains where \textsc{clean} finds model components, a process referred to as masking or boxing the source.  This step is usually done manually during the cleaning process. Unmasked or poorly masked images can cause the iterative model fitting done by \textsc{clean} to diverge, change the reconstructed flux of a source, and/or distort the properties of the noise (the so-called \textsc{clean} bias; \citealp{Condon1998TheSurvey}).

Manually masking images is an extremely time-consuming process, especially for the large data volumes currently produced by Atacama Large Millimeter/submillimeter Array (ALMA) and the National Science Foundation's Karl G. Jansky Very Large Array (VLA), and, in the future, by the Next Generation Very Large Array (ngVLA) and the Square Kilometer Array (SKA). It also hinders the production of automated, science-quality images via pipelines by limiting such pipelines to relatively shallow cleans that may not fully model the source. The latter is a particular concern for ALMA, where the science images and cubes typically have complex source structure \citep[e.g.,][]{Brogan2015TheRegion,Vlahakis2015The3.042}. A number of groups have developed automated masking algorithms, often in support of specific projects or surveys \citep{Cotton2007EVLAWindowing,Greisen2009AIPSAIPS, Kimball2011AutomatedCASA}. Pipelines for general purpose instruments like ALMA, however, require automated masking algorithms that can reliably mask a wide range of emission morphologies from simple point sources to rotating galaxies to protostellar outflows.

This paper describes a general purpose automated masking algorithm designed to mask emission with a wide range of emission morphologies while an interferometric image is being cleaned. We refer to this algorithm as \textsc{auto-multithresh} because it is an {\it auto}mated algorithm that uses {\it multi}ple {\it thresh}olds to mask emission. We describe  the \textsc{auto-multithresh} algorithm in general terms in Section~\ref{sec:description}. Section~\ref{sec:implementation} details the implementation of the algorithm in the CASA \citep{McMullin2007CASAApplications} task {\tt tclean}. The performance of the algorithm is characterized in Section~\ref{sec:performance}. We compare the \textsc{auto-multithresh} algorithm to a selection of other auto-masking algorithms in Section~\ref{sec:comparison}. We summarize our results in section~\ref{sec:summary}. 


\section{Description} \label{sec:description}

\begin{figure}
\centering
\includegraphics[height=8.0in]{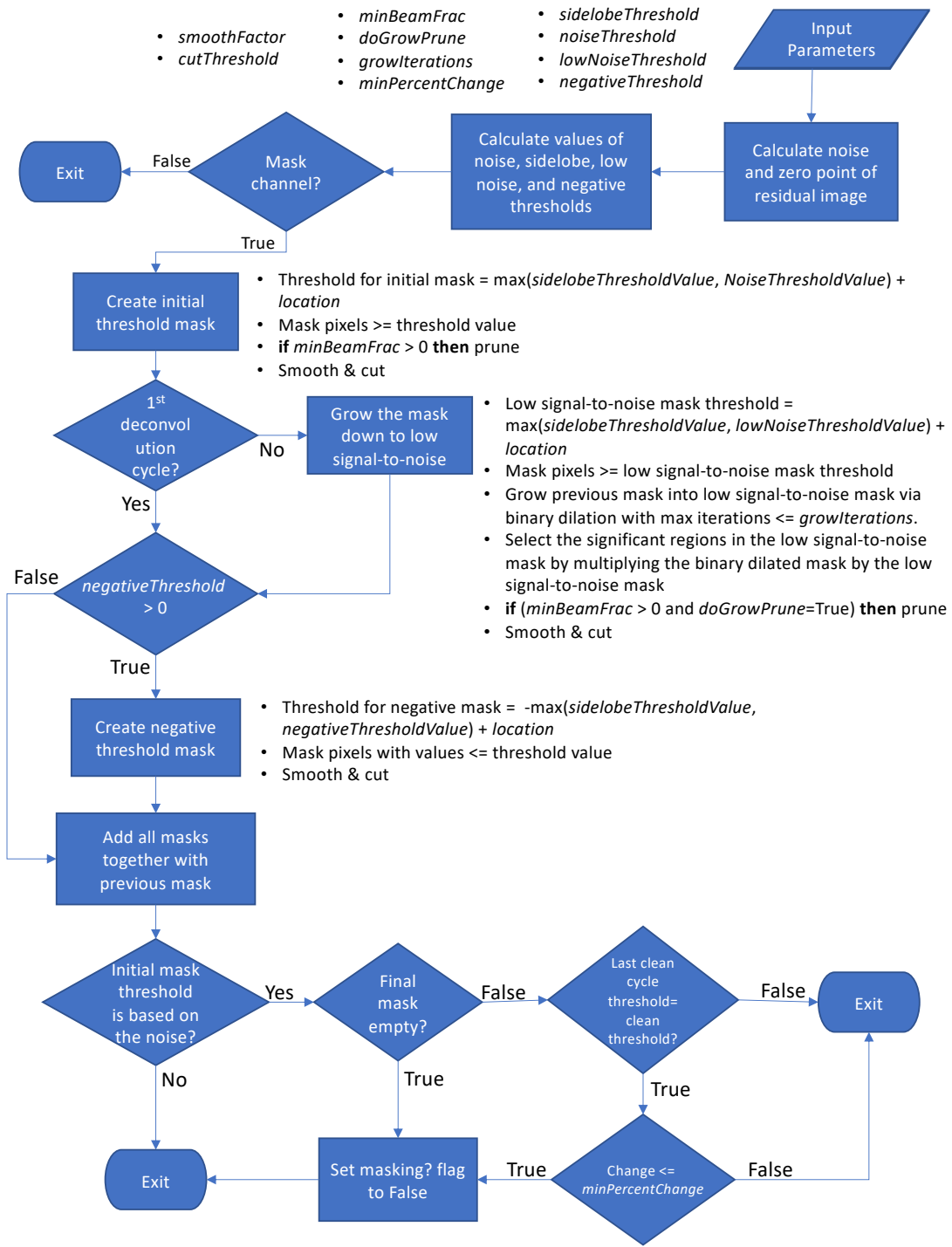}
\caption{A flowchart for the \textsc{auto-multithresh} algorithm. The input parameters are listed in the upper left. Details on a particular step in the algorithm are given to the right of the process. }
\label{fig:flowchart}
\end{figure}

We describe the \textsc{auto-multithresh} algorithm in detail below. The algorithm takes as its inspiration the peak finding steps from the \textsc{clumpfind} and \textsc{cprops} algorithms \citep{Williams1994DeterminingClouds,Rosolowsky2006BiasfreeProperties}, but extends them to work on the residual image within \textsc{clean}. Figure~\ref{fig:flowchart} provides a flowchart to illustrate the relationship between the different steps in the algorithm; the input parameters are defined in Table~\ref{tab:parameter_description}. To make the \textsc{auto-multithresh} algorithm as general as possible, all input parameters are defined using dimensionless values relative to the fundamental properties of the images (e.g., sidelobe level, signal-to-noise ratio, fraction of beam, etc) rather than using absolute numerical values. The exact numerical values corresponding to the input parameters are calculated during the masking process based on the image properties determined by \textsc{clean}.  The algorithm treats each channel in a cube independently to make the algorithm straightforward to parallelize.

\begin{table}
\caption{\textsc{auto-multithresh} Parameters} \label{tab:parameter_description}
{\sc Main Parameters}
\begin{description}
\item[{\it sidelobeThreshold}]  sets sidelobe level multiplier to consider  for the initial mask and the low signal-to-noise mask. Default: 3.0.
\item[{\it noiseThreshold}]  sets the signal-to-noise for the initial threshold mask. Default: 5.0.
\item[{\it lowNoiseThreshold}] sets the signal-to-noise  used for the low signal-to-noise mask. Default: 1.5.
\item[{\it negativeThreshold}] sets the signal-to-noise to use for creating a mask of negative (i.e., absorption) features. Default: 0.0.
\item[{\it minBeamFrac}] sets the minimum size of regions to be used in the mask as a fraction of the beam. Default: 0.3.
\end{description}
{\sc Secondary Parameters}
\begin{description}
\item[{\it fastNoise}] controls whether to use a simple median absolute deviation to calculate the noise (True) or the more complex procedure detailed in Appendix~\ref{sec:noise} (False). Default: True.
\item[{\it minPercentChange}] controls when to stop calculating a mask for a particular channel. Default: -1.0.
\item[{\it doGrowPrune}] controls whether to prune the low signal-to-noise mask. Default: True.
\item[{\it growIterations}] controls the maximum number of binary dilation iterations to use for the low signal-to-noise mask creation. Default: 75.
\item[{\it smoothFactor}] controls how to smooth the pixel masks. This parameter is usually left at its default value.  Default: 1.0.
\item[{\it cutThreshold}] controls how much of the smoothed pixel  mask to retain. This parameter is usually left at its default value. Default 0.01.
\end{description}
\end{table}

\subsection{\textsc{auto-multithresh} and the \textsc{clean} Algorithm}

The \textsc{auto-multithresh} algorithm is designed to work within \textsc{clean}, so that the derived mask can evolve as the model of the source is built up. Although a detailed discussion of \textsc{clean} is beyond the scope of this paper, we briefly outline the algorithm below to orient the reader. See the review by \citet{Rau2018SynergyAnalysis} and the references within for a more complete discussion. The \textsc{clean} algorithm consists of the following steps:
\begin{enumerate}
\item generate the residual visibilities by subtracting the model visibilities from the data visibilities. If no model components have been generated, the residual visibilities are the same as the data visibilities. \label{enum:resid_vis}
\item grid and inverse Fourier transform the residual visibilities to generate the residual image \label{enum:inv_ft}
\item deconvolve the image \label{enum:minor_cycle}
\begin{enumerate}
\item iteratively generate a model of the source by identifying emission peaks in the residual image \label{enum:model}
\item multiply the model components by a scaled PSF and subtract them from the residual image \label{enum:deconvolution}
\end{enumerate}
\item Fourier transform the current model image back into the visibility domain and de-grid the visibilities to generate new model visibilities \label{enum:ft}
\item subtract the model from the visibility data to update the residual visibilities \label{enum:subtract}
\item Repeat steps \ref{enum:inv_ft} through \ref{enum:subtract} until the residuals are below a threshold, a maximum number of model components have been generated, or the user stops the process. 
\end{enumerate}

Each iteration of the above process is collectively referred to as a major cycle. Typically a full run of clean\footnote{In this paper, when we talk directly about the algorithm we refer to it as \textsc{clean} , but when we talk about the process of producing an image we refer to it as cleaning. The latter is consistent with common usage in radio astronomy.} consists of several major cycles. Step~\ref{enum:minor_cycle} is commonly referred to as the minor cycle, or deconvolution step. During the minor cycle, components are added to the model until either it exceeds a set number of components per cycle or the residuals exceed a per-cycle threshold. 

The \textsc{auto-multithresh} algorithm operates on the residual image generated as the result of step~\ref{enum:inv_ft} at the start of the minor cycle. This allows the mask to include additional emission that is revealed as the brighter sources in the field are removed from the residual image.  

\subsection{Create Initial Threshold} \label{sec:initial_threshold}

The first stage of the \textsc{auto-multithresh} algorithm identifies significant emission peaks in the residual image. We refer to this stage as the initial thresholding. The key characteristic of the image for this stage is whether it is dominated by the sidelobes of the point spread function or noise. To determine this, we calculate two intermediate threshold values -- the sidelobe threshold and the noise threshold -- and use the values of these thresholds to determine the initial threshold.  Regions above the sidelobe threshold are unlikely to be the result of the point spread function of the observations, which is determined by the $u$-$v$ coverage of the observations and the weighting of the visibility data during gridding. The sidelobe threshold is calculated as:
\begin{equation}
sidelobeThresholdValue = sidelobeThreshold \times sidelobeLevel \times |peakResidual|
\end{equation}
where the {\it sidelobeLevel}  is the maximum of the absolute value of the sidelobe over all channels. It is calculated as 
\begin{equation}
    sidelobeLevel = \max ( |\min (PSF)|,  |\max(PSF - cleanBeam)|)
\end{equation}
where {\it cleanBeam} is the beam used by clean to restore the image. We take the absolute value of the {\it peakResidual} since it can be negative in images with strong absorption. The second intermediate threshold is the noise threshold. Regions above this threshold have signal-to-noise values that are consistent with significant emission. The noise threshold is calculated as:
\begin{equation}
noiseThresholdValue = noiseThreshold \times residualRMS
\end{equation}
Robustly estimating the {\it residualRMS} term in the above equation can be difficult because the residual image can include both noise and signal from the source, especially in the first few major cycles of clean. Appendix~\ref{sec:noise} details how we estimate the {\it residualRMS}. The threshold used for the initial threshold stage is then just the maximum of the two values, e.g.,
\begin{equation}
thresholdValue = \max(noiseThresholdValue, sidelobeThresholdValue) + location
\end{equation}
The {\it location} value adjusts the threshold to account for cases where the distribution of pixel values is centered around a non-zero value. The {\it location} was assumed to be zero in early versions of the algorithm, but has since been set to the median of the pixel values used in the noise calculation (see Appendix~\ref{sec:noise} for details).  Figure~\ref{fig:threshold} provides an example showing the initial thresholding step.

\begin{figure}
\centering
\includegraphics[width=\textwidth]{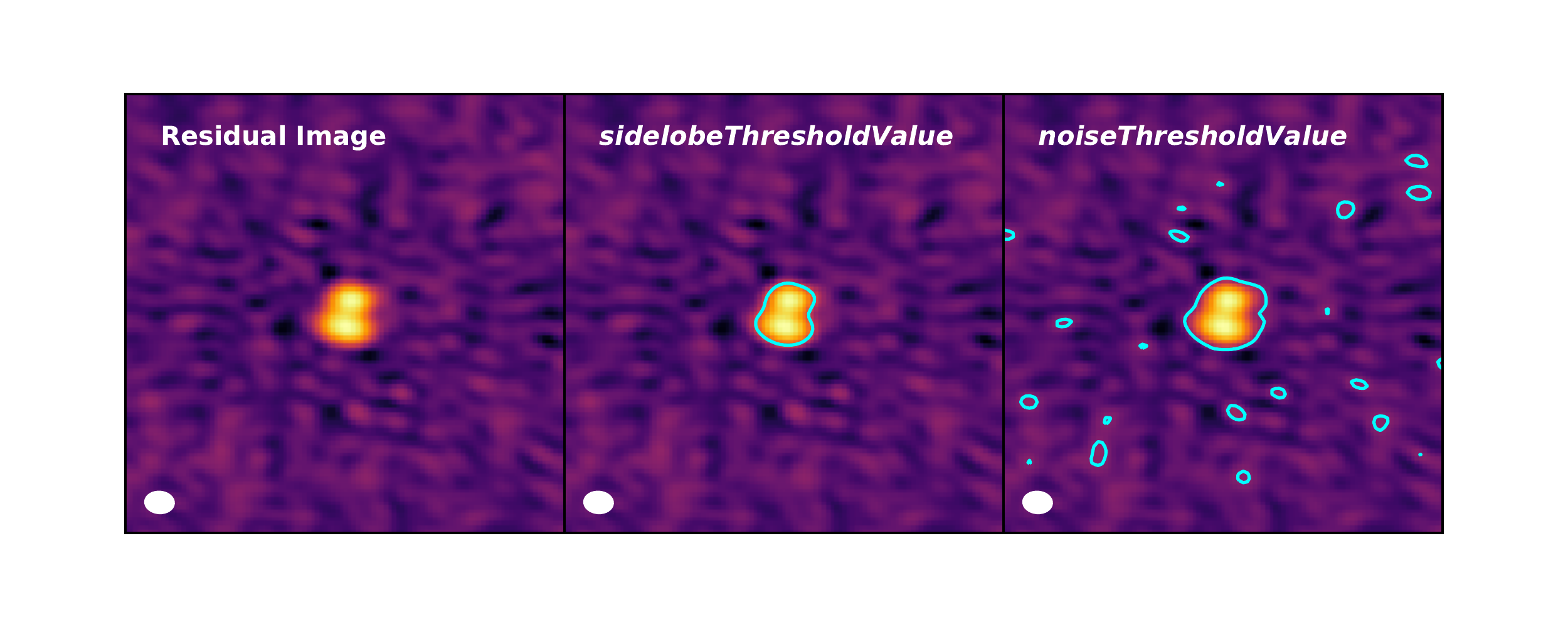}
\caption{Example of the initial thresholding. {\em Left:} Residual image with beam at bottom left. {\em Middle:} Residual image with contours indicating the intermediate mask based on the $sidelobeThresholdValue$. {\em Right:} Residual image with contours indicating the intermediate mask based on the $noiseThresholdValue$. The $noiseThresholdValue$ is lower than the $sidelobeThresholdValue$ and is clearly picking up features in the residuals due to sidelobes. Therefore, the $sidelobeThresholdValue$ was used to generate the initial threshold mask for this case.  Note that $noiseThreshold$ has been set here to a low value (1.75) for illustrative purposes.} \label{fig:threshold}
\end{figure}

\subsubsection{Pruning} \label{sec:prune}

\begin{figure}
\centering
\includegraphics[width=6.0in]{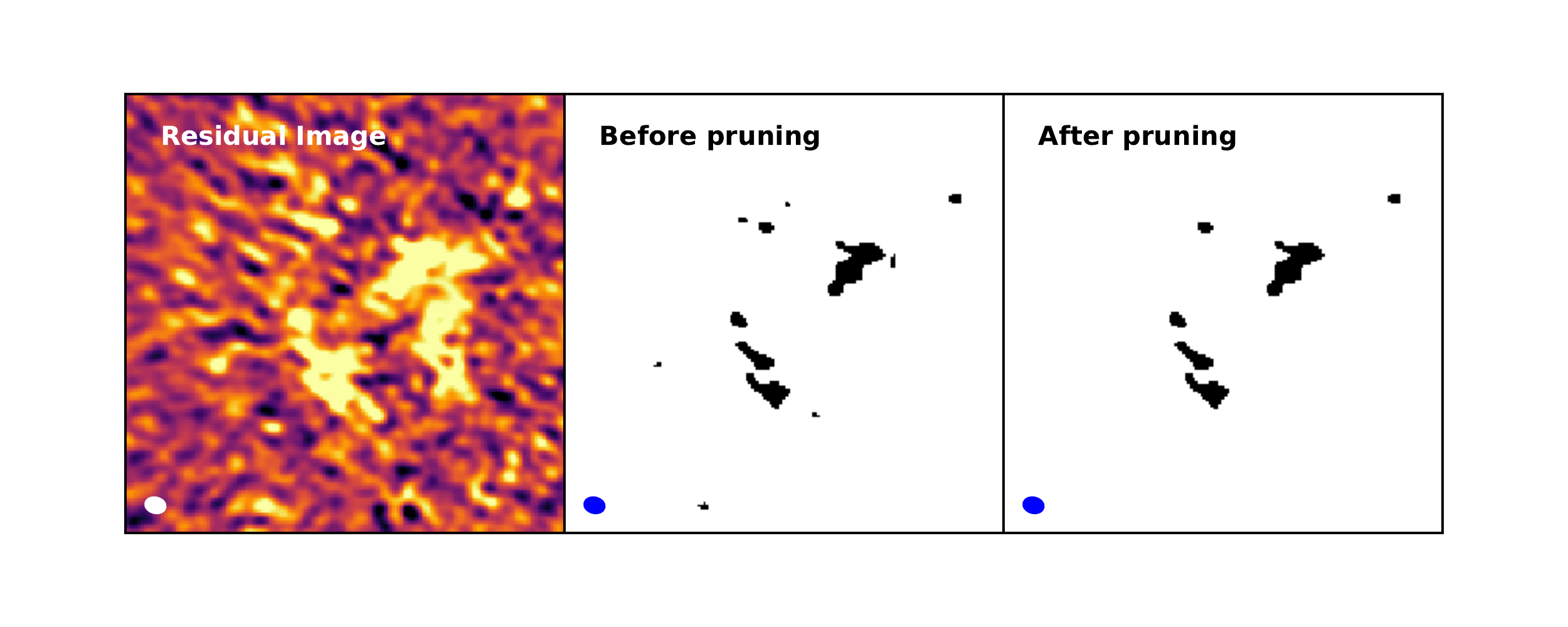}
\caption{A demonstration of pruning. {\it Left:} A residual image with the beam shown as a white ellipse in the lower left hand corner. {\it Middle:} Initial threshold mask generated from the residual image with masked pixels indicated in black and the beam shown in the lower left as a blue ellipse. Note the many regions of masked pixels that are smaller than the beam that appear to be associated with random noise peaks in the residual image. {\it Right:} Initial threshold mask after pruning regions with fewer pixels than 0.3 times the beam area (i.e.,  $minBeamFrac=0.3$). The beam is again shown in the lower left as a blue ellipse. Most of the small spurious regions have been removed.  The pruning is done on the initial threshold mask before it has been smoothed and cut to create a buffer around the masked regions. This is done so that the size of the region being removed is independent of the {\it smoothFactor} and {\it cutThreshold} parameters.} \label{fig:prune}
\end{figure}

The initial thresholding process often includes small, spurious regions that are more likely to due to the noise in the image rather than real emission. In general, one would like these regions pruned (i.e., removed) from the final mask to  avoid masking noise peaks, which can lead to poor models and, more seriously, divergence. To do this, patches of emission smaller than a fraction of the beam size are pruned from the mask via the following process
\begin{algorithmic}
\If{$npixelsInRegion < (minBeamFrac \times pixelsPerBeam)$}
\State{{\rm Remove region from mask}}
\EndIf
\end{algorithmic}
In general, the number of these small regions removed is roughly consistent with the number of false positives you would expect from a Gaussian noise distribution (when the {\it sidelobeThreshold} and {\it noiseThreshold} are set to appropriate values). Pruning can be turned off by setting {\it minBeamFrac} equal to zero. Figure~\ref{fig:prune} shows an initial threshold mask before and after pruning. Appendix~\ref{sec:prune_implementation} describes how pruning was implemented in the production version of \textsc{auto-multithresh}.

\subsubsection{Smoothing and Cutting the Mask} \label{sec:smoothandcut}

\begin{figure}
\centering
\includegraphics[width=\textwidth]{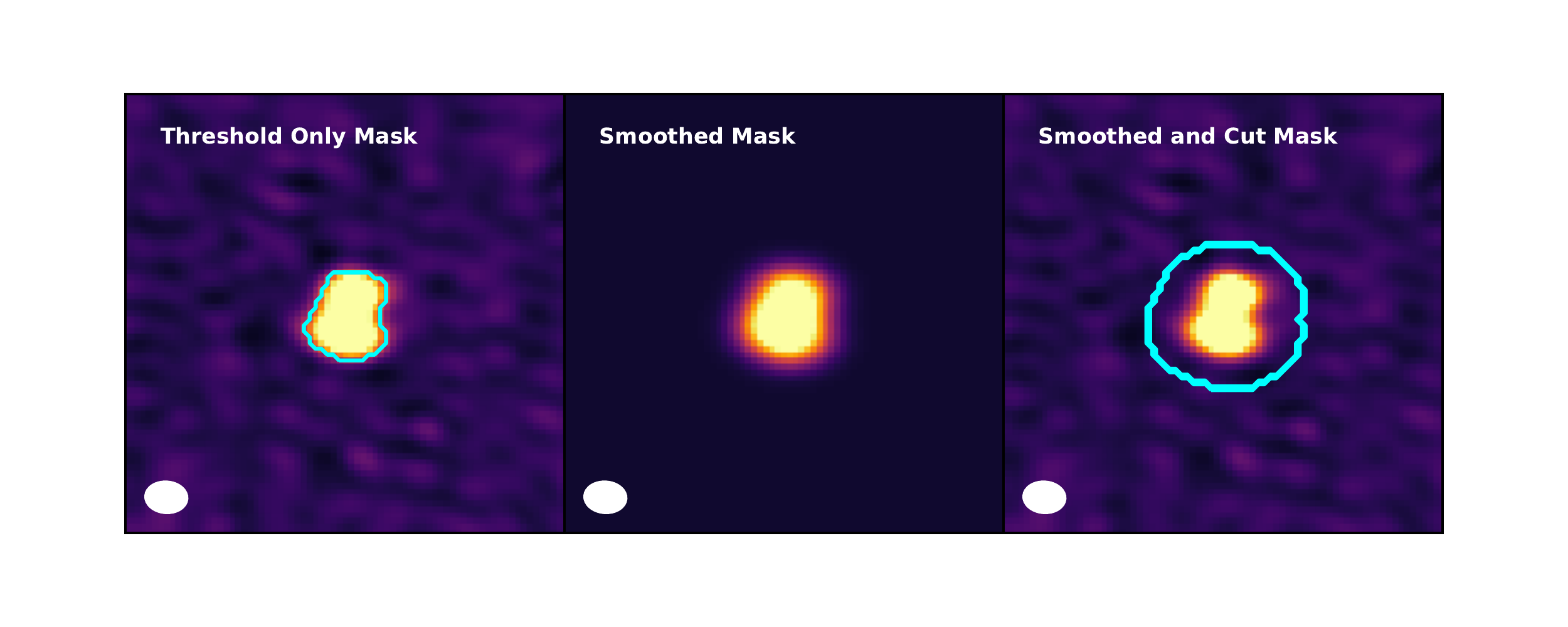}
\caption{Example of smoothing and cutting a mask. {\em Left:} Mask created after applying the initial threshold overlaid on the residual image. This mask leaves no margin between the bright emission and the mask. In this case, only the bright emission in the central mask would be cleaned; the fainter emission associated with it would be outside the mask and thus not cleaned. {\em Middle:} Colorscale image showing the mask from the left image smoothed by a Gaussian. In this case, the Gaussian was the size of the beam. {\em Right:} A mask created by retaining only the regions in the smoothed mask greater than 1\% of the peak. This mask has an appropriate margin between the bright emission and the edge of the mask that will allow any fainter emission associated with the peaks to be cleaned.}\label{fig:smoothcut}
\end{figure}

After pruning, the initial threshold mask is then expanded to include a buffer around the emission that has been identified as real. This step reflects common practice when manually imaging data to generously mask sources to avoid inadvertently excluding regions of more extended emission that often surround image peaks. The expansion is done by smoothing the initial threshold mask by a Gaussian that is a multiple of the beam size. The final threshold mask includes all regions above a fraction of the peak in the smoothed mask.
\begin{algorithmic}
\State $kernel = smoothFactor \times cleanBeam$
\State $convolvedMask = \textrm{convolve}(thresholdMask,kernel)$
\If{$convolvedMask > \max(convolvedMask) \times cutThreshold$}
\State set mask value to True
\EndIf
\end{algorithmic}
Figure~\ref{fig:smoothcut} demonstrates how the above operation expands the threshold mask to include a margin around the brightest emission. The parameters {\it smoothFactor} and {\it cutThreshold} are coupled. Increasing {\it smoothFactor} and {\it cutThreshold} by an appropriate amounts yields identical mask. In practice, we have found that setting the {\it smoothFactor} to 1.0 (i.e., the beam size) and {\it cutThreshold} to 0.01 (or 1\%) yields acceptable masks in all cases tested to date.

\subsection{Growing the Initial Threshold Mask Down to Lower Signal-to-Noise} \label{sec:grow}

\begin{figure}
\centering
\includegraphics[width=\textwidth]{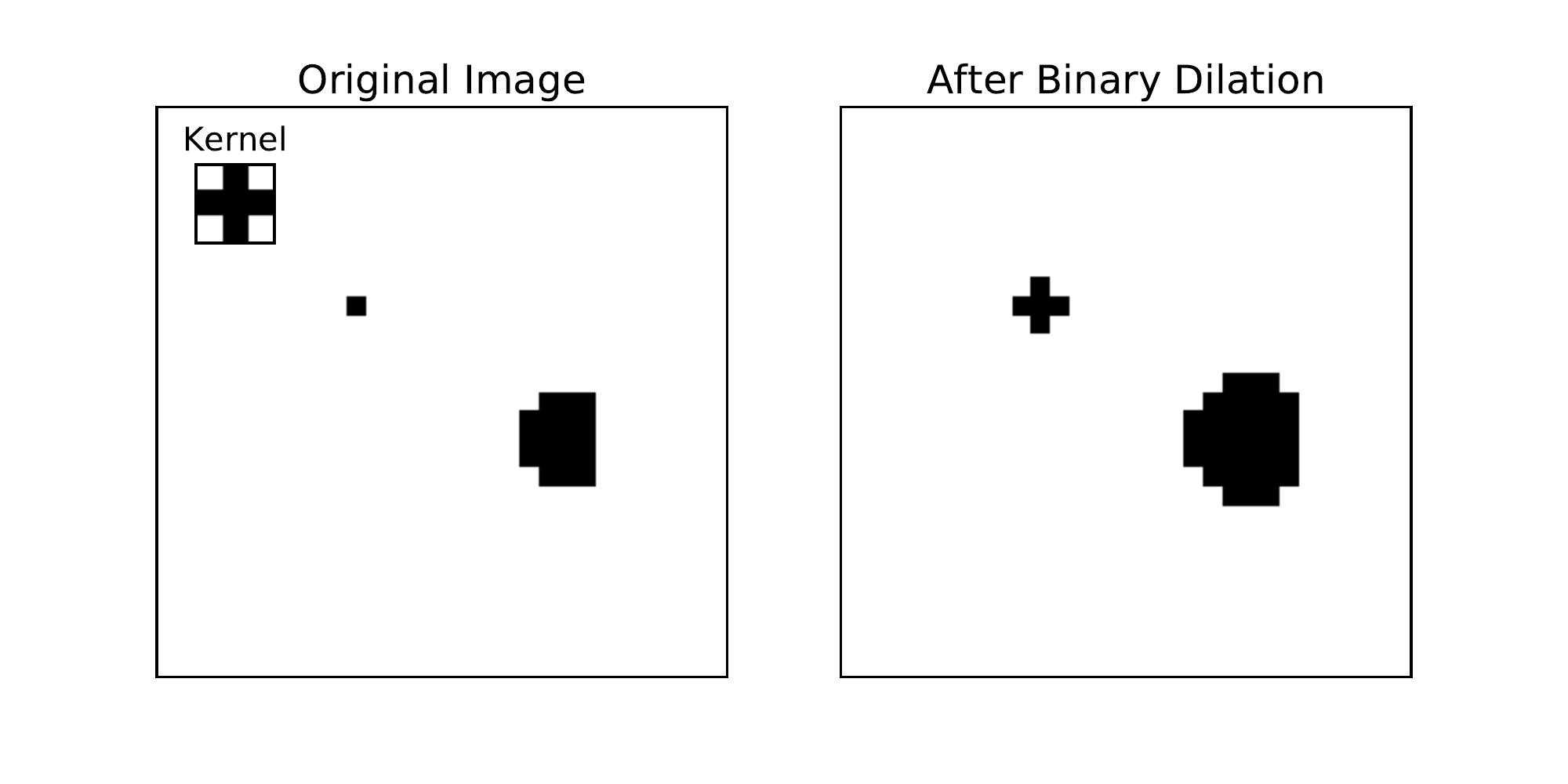}
\caption{Binary dilation example. {\em Left:} Original image with 3x3 cross kernel shown in upper left. {\em Right:} Resulting image after one iteration of binary dilation. Each positive pixel in the original image has been replaced by the kernel.} \label{fig:binary}
\end{figure}

\begin{figure}
\centering
\includegraphics[width=\textwidth]{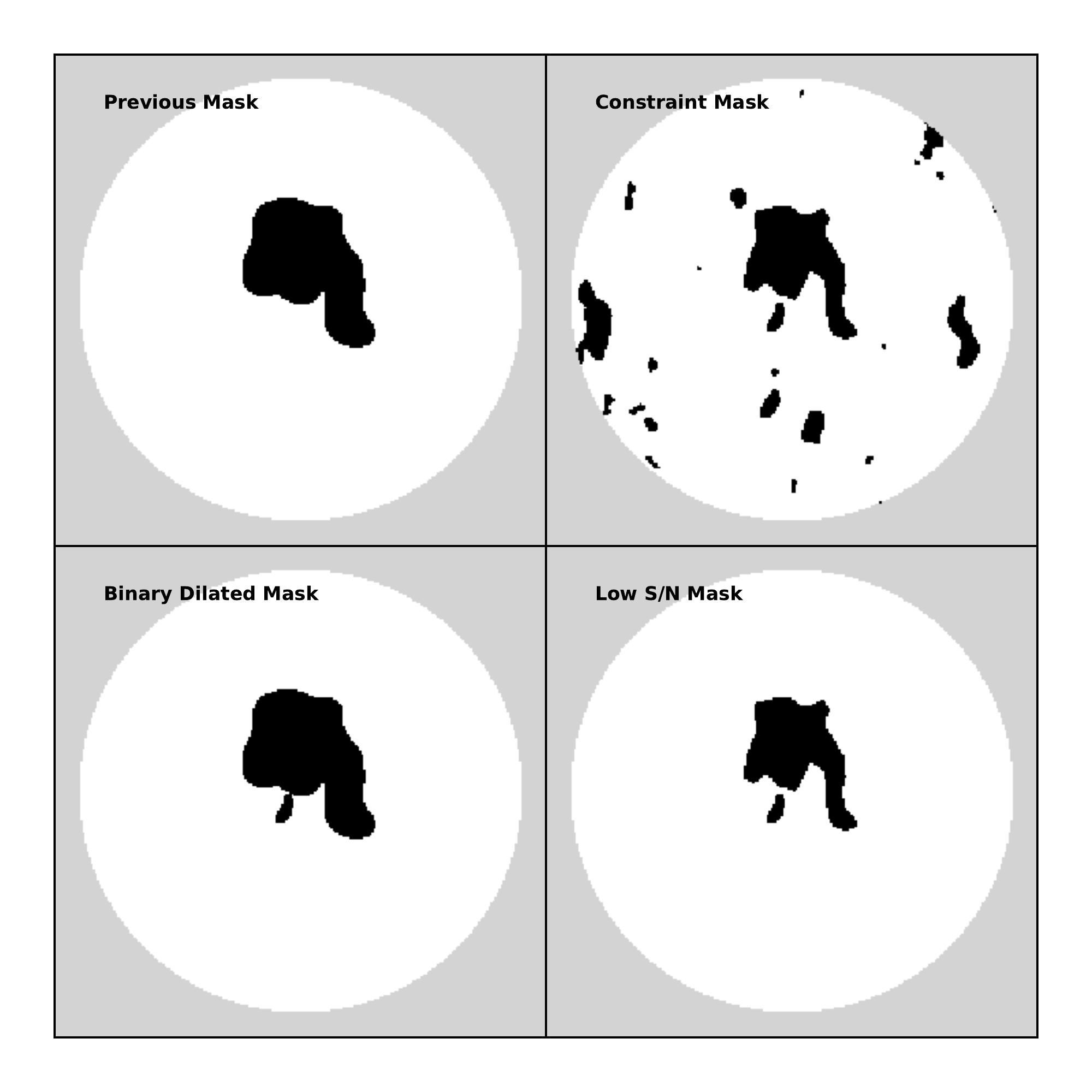}
\caption{Growing the initial threshold mask down to lower signal-to-noise. {\em Top Left:} Final mask from previous major cycle of clean. {\em Top Right:} Initial low signal-to-noise mask created by applying the $lowNoiseThreshold$ to the residual image. This mask is used as a constraint mask for the binary dilation step. {\em Bottom Left:} Mask created by binary dilating the previous clean mask into the low signal-to-noise mask. {\em Bottom Right:} Final low signal-to-noise mask created by multiplying the initial low signal-to-noise mask by the binary dilated mask to isolate only the low signal-to-noise regions associated with the previous clean mask. This mask is then pruned (optional), smoothed, and cut in the same fashion as the initial threshold mask. The gray regions in all panels are outside the primary beam mask.} \label{fig:grow}
\end{figure}

After the initial thresholding step, \textsc{auto-multithresh} then grows the initial threshold mask down to lower signal-to-noise to capture faint emission associated with significant emission peaks in the residual image. We refer to this mask as the low signal-to-noise mask. To expand the mask, we use a method called binary dilation -- an image processing technique analogous to convolution -- whereby each pixel in the mask is replaced by a kernel \citep[for more information on binary dilation, see][]{Russ2016TheHandbook}.  Figure~\ref{fig:binary} provides a simple example of the binary dilation process. The advantage of binary dilation over simple convolution is that the mask can be grown out to a constraint (i.e., boundary) by running the binary dilation multiple times with another mask as the constraint.  For \textsc{auto-multithresh}, we use a 3x3 cross kernel, which is the same kernel used in Figure~\ref{fig:binary}. The constraint mask is created using a similar process to the initial threshold, with the noise threshold set by the parameter {\it lowNoiseThreshold} 
\begin{algorithmic}
\State $lowNoiseThresholdValue = lowNoiseThreshold \times residualRMS $
\State $constraintMaskThreshold = \max(sidelobeThresholdValue,lowNoiseThresholdValue) + location$
\end{algorithmic}
The mask from the previous major cycle is iteratively expanded into the  constraint mask using binary dilation. The maximum number of iterations of binary dilation is controlled by the {\it growIterations} parameter. The constraint mask is then multiplied by the binary dilated mask to select only the low signal-to-noise regions associated with the previous mask. This process is done so that we can prune, smooth, and cut in this mask in the same fashion as the initial threshold mask.  Figure~\ref{fig:grow} demonstrates the entire process. The final result of this process is a mask that is been expanded (i.e., grown) into the low signal-to-noise regions surrounding the initial peaks. We refer to this mask as the ``grow'' or low signal-to-noise mask. This entire step is skipped if this is the first deconvolution cycle. To speed up masking, the pruning operation on the mask can be skipped by setting the parameter {\it doGrowPrune} to False.  In cases where negative emission is being masked, only the positive components of the previous mask are used in these steps.

\subsection{Create Negative Threshold Mask} \label{sec:negative}

Absorption is masked using a process similar to that used to create the initial threshold mask. The threshold level for the absorption mask is set via the user parameter {\it negativeThreshold}\footnote{The value for the {\it negativeThreshold} is assumed to be positive.}
\begin{algorithmic}
\State $negativeThresholdValue = negativeThreshold \times residualRMS $
\State $negativeMaskThreshold = -\max(negativeThresholdValue,sidelobeThresholdValue) + location$
\If{$residualImage <= negativeThresholdValue$}
\State set mask value to True
\EndIf
\end{algorithmic}
The subsequent mask is smoothed and cut using the same process as described above in Section~\ref{sec:smoothandcut} for the initial threshold mask. This step is skipped if negative threshold is set to zero. The negative threshold mask is not pruned and is not cascaded down to lower signal-to-noise levels. While this is possible in theory, in practice this would significantly slow down process of masking. Therefore, in general, absorption masking is less complete than emission masking. The absorption masks are tracked separately from the emission masks since these emission components are generally distinct (although related) features.

\subsection{Combine Masks}

The final mask is created by combining the initial threshold mask, the grow mask, and the absorption mask with the mask from the previous clean cycle (if present). This combined mask is then used in the subsequent minor (deconvolution) cycle.

\subsection{Control Future Masking}

At the end of the masking process, we determine whether or not further masking is warranted in future major cycles for a particular channel. Generating the masks is a computationally expensive process that in many cases does not need to be repeated for all channels for every major cycle. An internal variable called {\it chanflag} is used to track whether or not a channel should be masked. A value of False tells \textsc{auto-multithresh} to mask that channel; a value of True tells \textsc{auto-multithresh} to skip masking that channel. The value for all channels is initially set to False (mask each channel). The value for the variable is re-evaluated at the end of the masking cycle and set to True if the noise threshold is being used for that channel and any of the following conditions are true:
\begin{itemize}
\item the final mask is empty
\item the previous clean cycle had the cycle threshold equal to the clean threshold and the change in the mask was less than user parameter {\it minPercentChange}, where {\it minPercentChange} is defined as $100 * ( n_{pix,current} - n_{pix,previous} ) / n_{pix,previous}$
\end{itemize}
The first condition is always enabled. The second condition is turned off ($minPercentChange=-1.0$) by default. If enabled, it can significantly reduce total clean time  when set to a conservative value (1-2\%) since the mask will stop updating at some point during the clean. However, this speed-up can come at the expense of undermasking in cubes which have many channels of similarly bright complex emission. This behavior occurs because the cycle threshold can reach the clean threshold multiple times during a complex clean. If the mask is stopped at the first time the cycle threshold is equal to the clean threshold, then the masking cannot adapt to the subsequent changes in the residuals from future masking cycles.



\section{Development and Implementation} \label{sec:implementation}

The end goal of the \textsc{auto-multithresh} algorithm  was to enable deeper cleaning of images in the ALMA Imaging Pipeline. Prior to Cycle 5, the images were cleaned shallowly down to four times the noise multiplied by a dynamic range correction with a simple primary beam mask. Deeper cleaning required better masking of the complex emission typically seen in ALMA images and cubes. To develop and test potential algorithms to do this, we used the Cycle 5 benchmark suites to provide a set of representative ALMA data. The Cycle 5 benchmark included approximately 45 data sets from ALMA PI projects from previous cycles that were successfully run on the telescope and used to test both calibration and imaging in the ALMA Pipeline. One of the selection criteria for these projects was that they span a range of imaging cases typically seen by ALMA, including point sources, extended objects both with and without missing flux, and absorption. For each data set, the ALMA Imaging Pipeline produces four to five continuum images and multiple line cubes, resulting in over 1000 test images total. This large test corpus enabled us to ensure that our algorithm would work in general for the wide range of data that is processed by the ALMA Pipeline. It also allowed us to remove the quality of the calibration as a variable since all these projects had been calibrated by the ALMA Pipeline using the best practices for ALMA data. As the calibration produced by the ALMA Pipeline evolved, updated calibrations for the data were obtained by re-running a newer version.

The algorithm was prototyped in Python so that we could rapidly explore and compare different methods for masking emission \citep{Kepley2019Auto-multithresh:Clean}. The prototype used {\it PySynthesisImager}, which is a wrapper Python class within CASA built on top of the collection of synthesis imaging Python tools. Each of these tools has a one-to-one mapping of C++ classes and methods. The CASA {\tt tclean} task is also built on top of {\it PySynthesisImager}. The masking code was inserted prior to the minor cycle. A small subset of the continuum images were used to develop the initial algorithm. Initial tests were largely done in CASA 5.0 pre-release builds. 

\begin{figure}
    \centering
    \includegraphics[width=\textwidth]{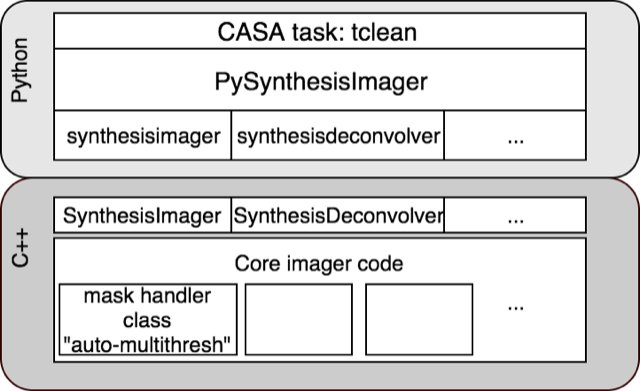}
    \caption{Overview of \textsc{auto-multithresh} implementation within the CASA {\tt tclean} task. There is a one-to-one correspondence between the Python $PySynthesisImager$ and C++ classes and methods. The \textsc{auto-multithresh} algorithm is implemented within the mask handler class in the C++ layer.}
    \label{fig:auto_multithresh_implementation}
\end{figure}

The final algorithm was then integrated within the CASA {\tt tclean} task for speed and better integration with other imaging features and performance. It was implemented as part of the mask handling C++ module within the refactored imager code and launched from the deconvolver to be in sync with its iteration control as illustrated in Figure~\ref{fig:auto_multithresh_implementation}. Early versions of the algorithm were present in CASA 4.7.1 and 5.0.0. CASA 5.1.1  included the first fully validated implementation of the algorithm. It was tested on the full set of of $\sim1000$ benchmark images used to validate the ALMA Cycle 5 Pipeline. The \textsc{auto-multithresh} algorithm has been used in production as part of the ALMA Imaging Pipeline since Cycle 5. 

Based on experience with the algorithm in production in ALMA Cycle 5, significant speed improvements were made in CASA version 5.3, including
\begin{itemize}
\item under-the-hood speed improvements to the pruning code, 
\item dynamically determining the number of iterations used in the grow step rather than using a fixed value,
\item ability to turn off the prune in the grow step, and
\item logic to determine whether the masking calculations needed to be repeated for a channel in the future
\end{itemize}
The final version of this algorithm as described in this paper is in CASA version 5.5 and beyond. It introduced improvements to the estimation of the noise in the residual image and included the location of the distribution in the threshold calculations.


\section{Performance} \label{sec:performance}

This section details the performance of the \textsc{auto-multithresh} algorithm, in terms of mask quality (\S\ref{sec:quality}), sensitivity to input parameters (\S\ref{sec:input_params}), and speed of the algorithm (\S\ref{sec:speed}). 

\subsection{Mask Quality} \label{sec:quality}

The ultimate goal of masking emission during cleaning is to produce a model that accurately reflects the true sky brightness distribution.  In general, a good mask is defined as one that it encompasses all the significant emission in an image, while excluding obvious artifacts. Common practice generally leaves a beam size buffer between the emission regions and the edge of the mask to avoid constraining the clean too tightly. 

The quality of the masks produced by \textsc{auto-multithresh} was evaluated by eye by experts at imaging interferometric data. The initial evaluation of the mask quality on a small subset of ALMA images was done by the authors of this paper as part of the development of the prototype algorithm. To validate the use of \textsc{auto-multithresh} in the ALMA Cycle 5 Pipeline, the first author (A. Kepley) examined all the masks produced by the ALMA Cycle 5 benchmark data sets, which included approximately 1000 images. Co-authors C. Brogan and R. Indebetouw also examined a significant fraction of these images.

These initial expert evaluations of the quality of the masks produced by \textsc{auto-multithresh} have been validated by its performance in production as part of the ALMA Imaging Pipeline, starting in ALMA Cycle 5.  Approximately 70\% of all ALMA project in Cycle 5 and 95\% of all ALMA project in Cycle 6 were processed by the ALMA Pipeline. All resulting data sets are inspected by hand by experienced data reducers to determine whether they meet the science goals for the proposal. This inspection includes an evaluation of the masks and the final cleaned images. Any masking deficiencies are reported to the first author. In general, we have received very few (less than 25) reports of issues over the two full cycles that \textsc{auto-multithresh} has been used in production as part of the ALMA Pipeline. The most difficult cases for the algorithm are those with extended emission covering a large fraction of the field of view and/or with strong absorption. These cases were  ameliorated by the improved noise estimate described in Appendix~\ref{sec:noise}.

Figures~\ref{fig:ngc4414} and \ref{fig:rcw120} show examples of two of the more complex images and their masks generated by the ALMA Cycle 6 Imaging Pipeline with the default parameters given in Table~\ref{tab:pipeline_params} and no human intervention. Beyond its use in the ALMA Imaging Pipeline, \textsc{auto-multithresh} has also been used to successfully mask images from the Jansky Very Large Array (JVLA) and the Australia Telescope Compact Array (ATCA). For examples,  see Petter et al. (in prep) and \citealt{Wenger2019TheCatalog}. 



\begin{figure}
    \centering
    \includegraphics[width=\textwidth]{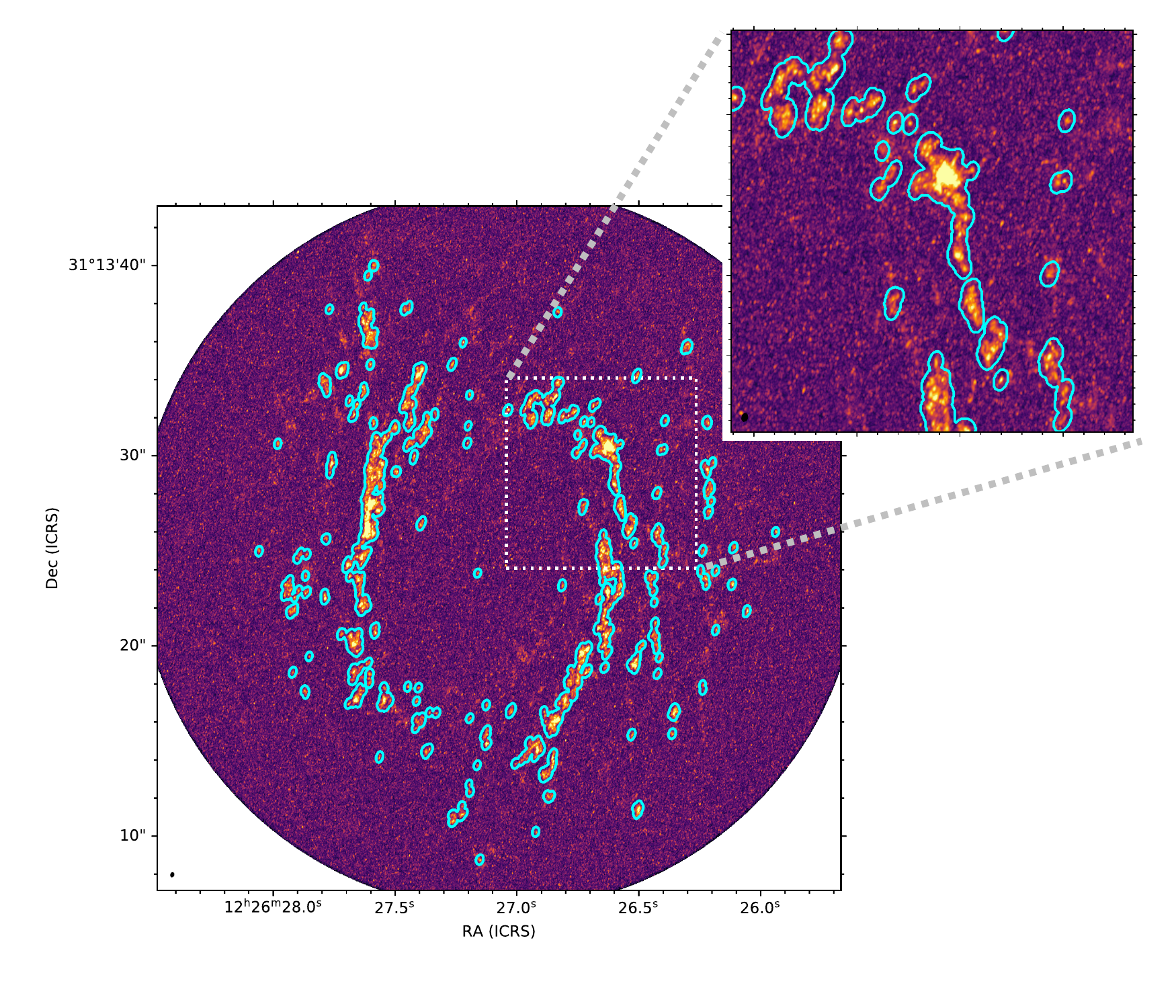}
    \caption{Peak intensity image of \co{12}{2}{1} emission from the face on galaxy NGC4414 showing emission from the entire disk as well as a zoom-in on a typical outer region of the galaxy. A flattened version of the per-channel mask produced by \textsc{auto-multithresh} is shown as cyan contours. The mask was produced by \textsc{auto-multithresh} without human intervention using the fiducial parameters given in Table~\ref{tab:pipeline_params}. The mask captures the complex emission in the region. The beam is shown in the lower left of the main image and the inset image.}
    \label{fig:ngc4414}
\end{figure}

\begin{figure}
    \centering
    \includegraphics[width=\textwidth]{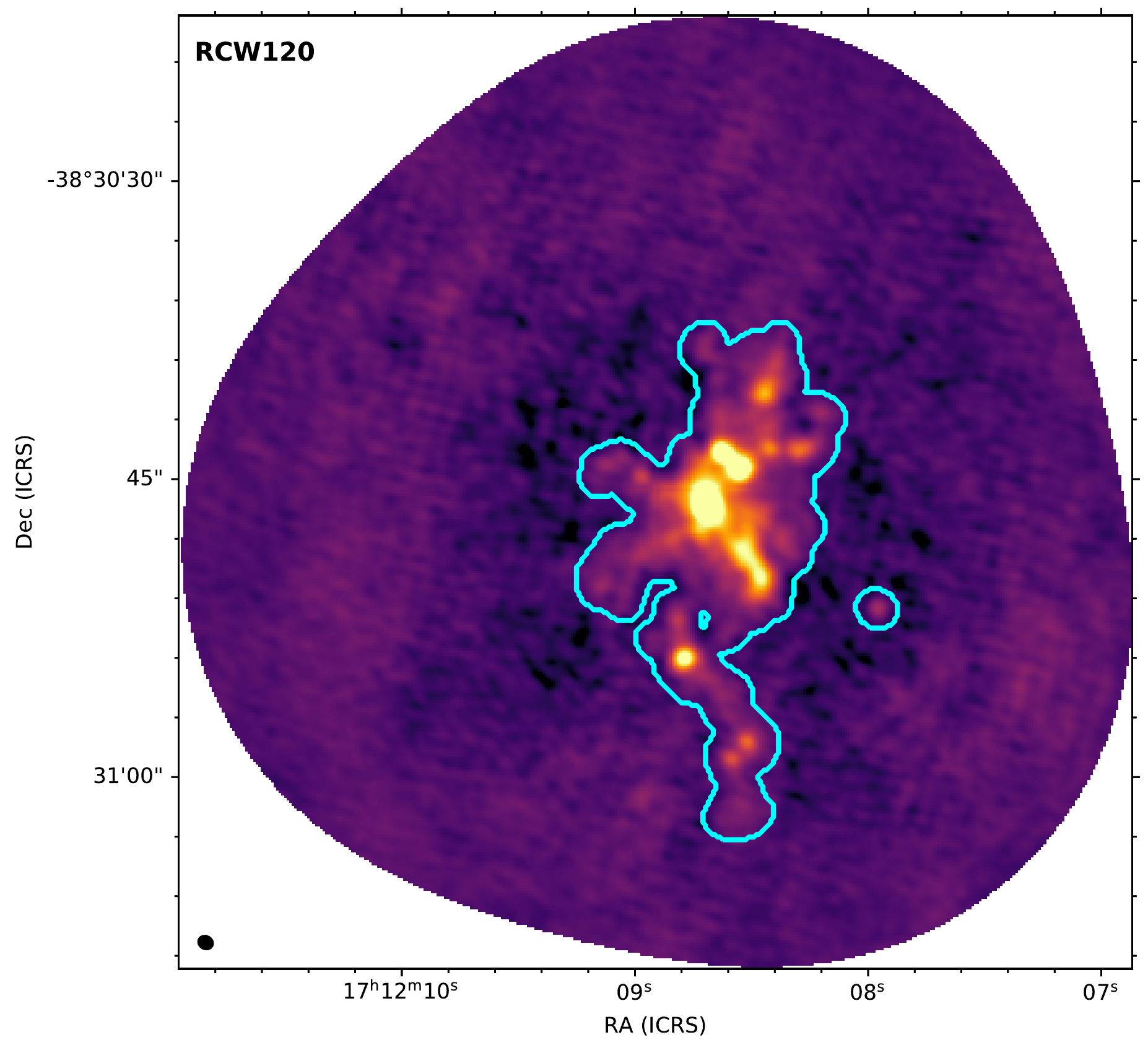}
    \caption{A 223.6GHz continuum image of the HII region RCW120. The mask produced by \textsc{auto-multithresh} without human intervention using the fidicual parameters given in Table~\ref{tab:pipeline_params} is shown as cyan contours. The beam is shown in the lower left of the image.}
    \label{fig:rcw120}
\end{figure}

\subsection{Input Parameter Sensitivity} \label{sec:input_params}

\begin{figure}
\centering
\includegraphics[width=\textwidth]{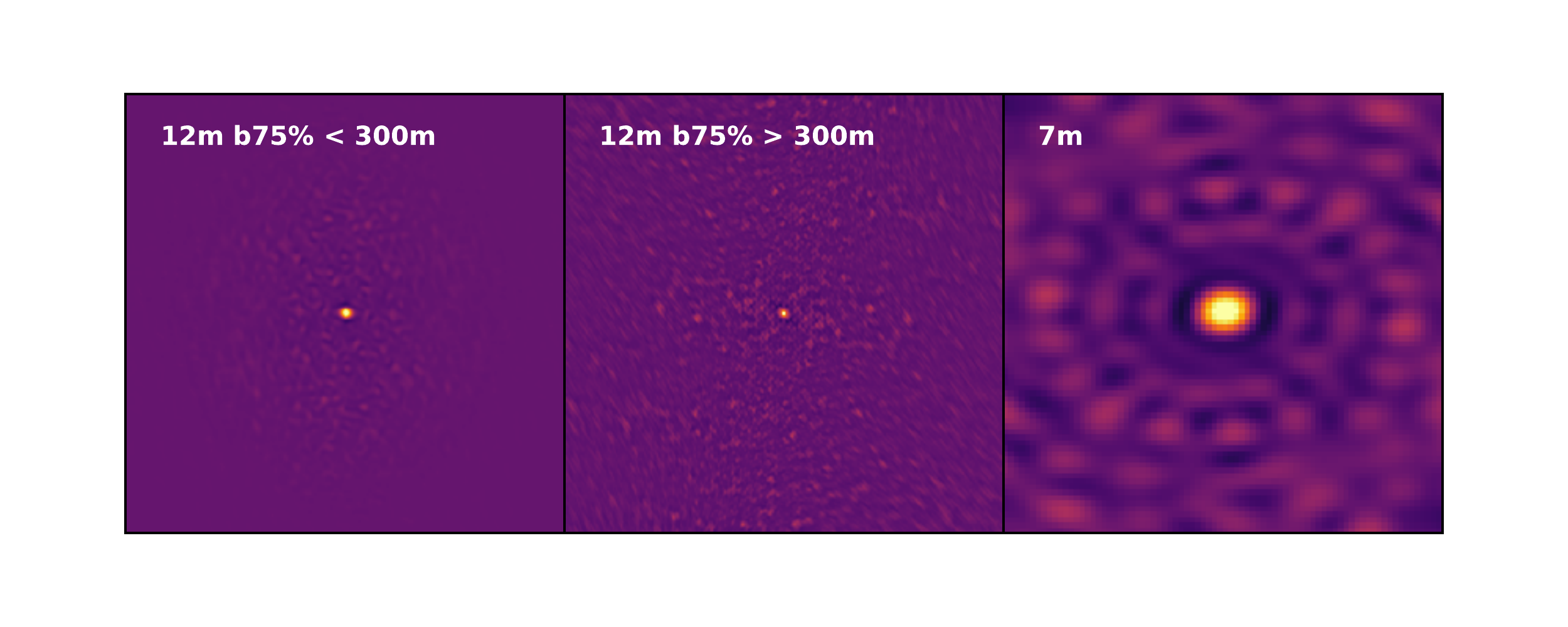}
\caption{The optimal \textsc{auto-multithresh} parameters for an image show a moderate dependence on the quality of its point spread function (i.e., {\it u-v} coverage). The ALMA Cycle 5 and 6 Imaging Pipelines split images into three broad categories and use a slightly different set of parameters for each. The parameters are detailed in Table~\ref{tab:pipeline_params}. Here we show example point spread functions for the three different sets of parameters with the same logarithmic stretch. {\em Left:} The point spread function for compact (75 percentile baseline less than 300m) ALMA 12m configurations. This point spread function has very low sidelobes. {\em Middle:} The point spread function for extended (75 percentile baseline greater than 300m) ALMA 12m configurations. This point spread function has slightly higher sidelobes than the compact ALMA configuration. {\em Right:} The point spread function for the ALMA 7m array. Note the extremely high sidelobes. } \label{fig:psf_example}
\end{figure}

The \textsc{auto-multithresh} user parameters are specified relative to the fundamental properties of the images, so that the same set of parameters can be used for different images. Based on our experience with the ALMA Imaging Pipeline, we have found that the parameters are sensitive to the quality of the {\it u-v} coverage of the observations, which corresponds to the point spread function (PSF) of the image. Table~\ref{tab:pipeline_params} shows the three different sets of parameters used for the ALMA Pipeline for short baseline 12m data (best {\it u-v} coverage), long baseline 12m data (average {\it u-v} coverage), 7m data (poor {\it u-v} coverage). A typical PSF for each case is shown in Figure~\ref{fig:psf_example}. The main parameters responsible for the quality of the mask are the {\it sidelobethreshold}, {\it noiseThreshold}, {\it lowNoiseThreshold}, {\it negativeThreshold}, and {\it minBeamFrac}.\footnote{The {\it growIterations} parameter value was decreased for 12m cubes solely reduce extremely long run times in production for a subset of cubes. For most cases, reducing this parameter does not affect the quality of the masks because the number of binary dilation iterations is typically much less than 50, except in extremely complex cases.}   Tests with the ALMA Cycle 6 Pipeline suggest that changing the Briggs robust weighting parameters do not appear to affect the masks significantly, even though these parameters change the PSF by changing the weighting of the individual visibilities in the gridding step of \textsc{clean}. 

When \textsc{auto-multithresh} is used with data from different telescopes, the user parameters have to be adjusted for the different {\it u-v} coverages of the different instruments. For VLA observations with good {\it u-v} coverage, the parameters are generally very similar to those for short baseline ALMA 12m data, which has similarly good {\it u-v} coverage. Data from the Australian Telescope Compact Array (ATCA) required significant tweaking of the \textsc{auto-multithresh} parameters because it has worse {\it u-v} coverage than the ALMA 12m array and the VLA \citep{Wenger2019TheCatalog}. We present detailed guidelines for tuning the \textsc{auto-multithresh} parameters in Appendix~\ref{sec:tuning}.

\begin{deluxetable}{lcccccccc}
\tabletypesize{\scriptsize}
\tablewidth{\textwidth}
\tablecolumns{9}
\tablehead{
\colhead{} & 
\colhead{75\% percentile } & 
\colhead{Image} &
\colhead{} &
\colhead{} &
\colhead{} &
\colhead{} & 
\colhead{} &
\colhead{} \\
\colhead{array} & 
\colhead{baseline (m) } &
\colhead{Type} &
\colhead{{\it sidelobeThreshold}} &
\colhead{{\it noiseThreshold}} &
\colhead{{\it lowNoiseThreshold}} &
\colhead{{\it negativeThreshold}} & 
\colhead{{\it minBeamFrac}} &
\colhead{{\it growIterations}\tablenotemark{a}}
}
\tablecaption{ Cycle 5 and 6 `auto-multithresh' parameters \label{tab:pipeline_params}}
\startdata
12m & $<300$ & continuum & 2.0 & 4.25 & 1.5 & 0.0 & 0.3 & 75 \\
12m & $<300$ & line & 2.0 & 4.25 & 1.5 & 15.0 & 0.3 & 50 \\  \hline
12m & $>300$ & continuum & 3.0 & 5.0 & 1.5 & 0.0 & 0.3 & 75 \\
12m & $>300$ & line & 3.0 & 5.0 & 1.5 & 7.0 & 0.3 & 50 \\ \hline
7m & \nodata & line \& continuum & 1.25 & 5.0 & 2.0 & 0.0 & 0.1 & 75  \\
\enddata
\tablenotetext{a}{The {\it growIterations} parameter value was decreased for 12m cubes to reduce extremely long run times in production for a sub-set of cubes. For most cases, reducing this parameter does not decrease the quality of the masks because the number of binary dilation iterations is typically much less than 50, except in extremely complicated cases.}
\end{deluxetable}

\begin{figure}
    \centering
    \includegraphics[width=\textwidth]{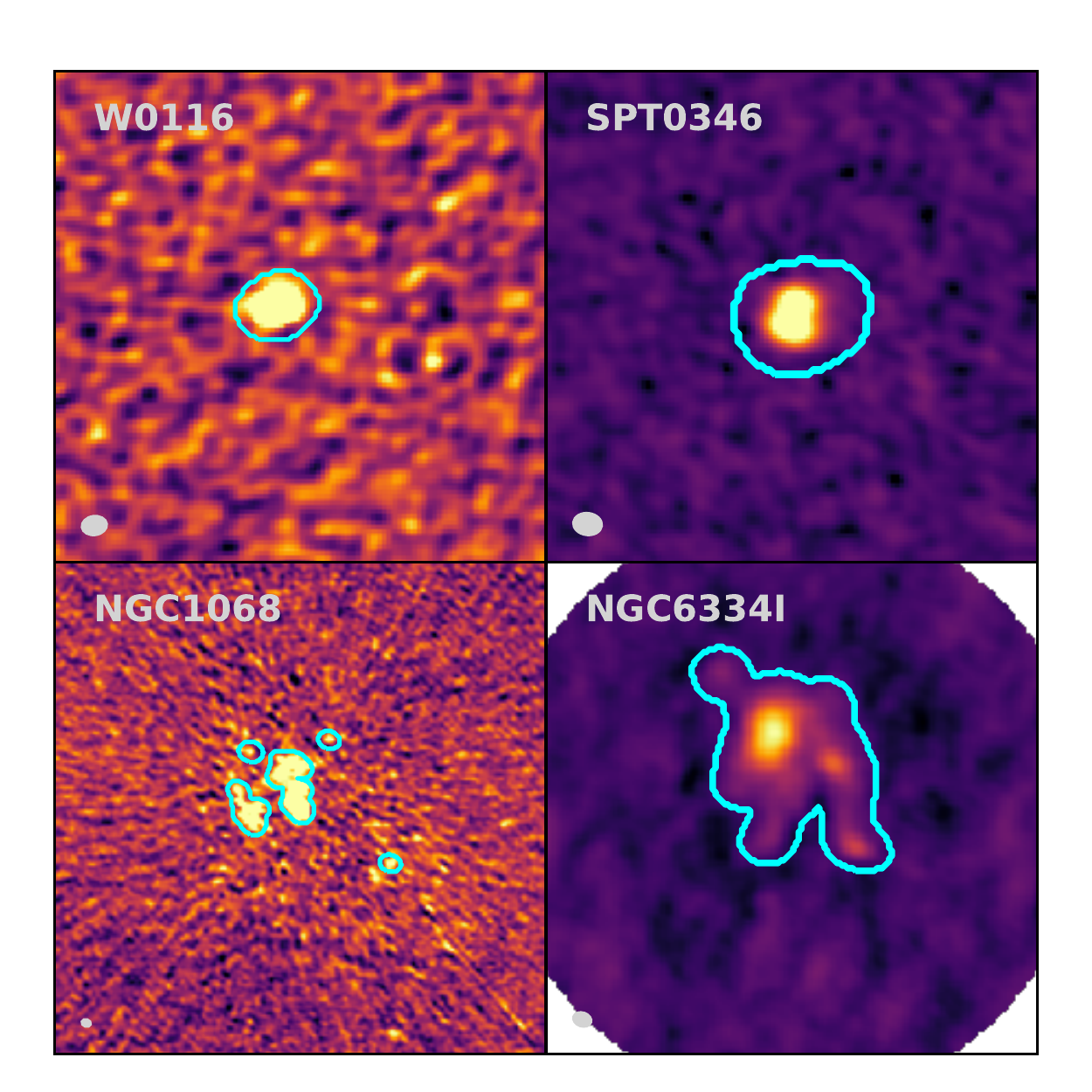}
    \caption{Images used to explore how varying the \textsc{auto-multithresh} parameters changes the resulting mask. These ALMA images were chosen to span a wide range of imaging cases. {\em Top Left:} The W0116 image is dominated by noise. {\em Top Right:} The SPT0346 image is dominated by sidelobes. {\em Bottom Left:} The NGC1068 image has complex emission as well as significant artifacts due to the underlying issues with the data. {\em Bottom Right:} The NGC6334I image has significant low signal-to-noise emission. For all images, the resulting mask from the fidicual \textsc{auto-multithresh} parameters ($sidelobeThreshold=3.0$, $noiseThreshold=5.0$, $lowNoiseThreshold=1.5$) is shown in cyan. The beams are shown in gray in the lower left hand corner of each panel.} 
    \label{fig:mask_example}
\end{figure}

To demonstrate the effects of modifying the various \textsc{auto-multithresh} parameters, we use four example data sets shown in Figure~\ref{fig:mask_example}. The data sets were selected to span a wide range of cases: the W0116 image is dominated by noise, the SPT0346 data set is dominated by sidelobes, the NGC1068 data set has complex emission as well as significant artifacts due to issues with the underlying data, and NGC6334I has large amounts of low signal-to-noise emission. The control images in Figure~\ref{fig:mask_example} were produced in CASA 5.3.0-143 using the following parameter values for all the images: 
\begin{itemize}
    \item $sidelobeThreshold = 3.0$, 
    \item $noiseThreshold = 5.0$, 
    \item $lowNoiseThreshold = 1.5$, 
    \item $minBeamFrac = 0.3$, 
    \item $negativeThreshold = 0.0$.
    \item $growIterations=75$
    \item $minbeamfrac=0.3$
\end{itemize}
They were cleaned down to a  signal-to-noise of 2.0 multiplied by a dynamic range modifier determined by the ALMA Imaging Pipeline based on the image properties. After creating the initial images, we varied one of three main parameters ({\it sidelobeThreshold}, {\it noiseThreshold}, and {\it lowNoiseThreshold}) while holding the others constant.

Table~\ref{tab:parameter_variation} shows the results of these tests. We list the number of pixels in each mask, maximum, minimum,  mean, and noise (estimated using the scaled median absolute deviation) in the final image as well as the difference between these quantities for the control and test images.  This table highlights the importance of the {\it sidelobeThreshold} in the mask determination; changing the value of this parameter significantly changes the mask in all cases. The reason for this influence is that the {\it sidelobeThreshold} plays a role in both the initial threshold mask and the low signal-to-noise threshold mask. In general, for images with significant emission, the initial threshold mask is set by the {\it sidelobeThreshold} for  the first few major cycles, then is set by the {\it noiseThreshold} for what remains of the clean. After the first major cycle, the initial threshold mask is also cascaded down to lower signal-to-noise using a threshold set by the maximum of the  {\it lowNoiseThresholdValue} or the {\it sidelobeThresholdValue}. This cascade is often set by the {\it sidelobeThreshold} rather than the {\it lowNoiseThreshold}.

\begin{deluxetable}{lrrrrrrrrrr}
\tablewidth{0pt}
\tabletypesize{\scriptsize}
\tablecaption{Image Properties for Different \textsc{auto-multithresh} Parameters \label{tab:parameter_variation}}
\tablecolumns{11}
\tablehead{
        \colhead{} &
        \colhead{} &
        \colhead{} &
        \colhead{Maximum} &
        \colhead{} &
        \colhead{Minimum} &
        \colhead{} &
        \colhead{Mean} &
        \colhead{} &
        \colhead{Noise} &
        \colhead{} \\
        \colhead{Image} &
        \colhead{$N_{pix,mask}$} &
        \colhead{\%$\Delta$} &
        \colhead{Jy~beam$^{-1}$} &
        \colhead{\%$\Delta$} &
        \colhead{Jy~beam$^{-1}$} &
        \colhead{\%$\Delta$} &
        \colhead{Jy~beam$^{-1}$} &
        \colhead{\%$\Delta$} &
        \colhead{Jy~beam$^{-1}$} &
        \colhead{\%$\Delta$} \\
        \colhead{(1)} &
        \colhead{(2)} & 
        \colhead{(3)} &
        \colhead{(4)} &
        \colhead{(5)} &
        \colhead{(6)} &
        \colhead{(7)} & 
        \colhead{(8)} & 
        \colhead{(9)} & 
        \colhead{(10)} & 
        \colhead{(11)} 
}
\startdata
\cutinhead{ SPT0346 } 
 Control &   687 & \nodata & 4.622e-03 & \nodata & -1.568e-04 &  \nodata  & 1.333e-05 & \nodata & 4.004e-05 & \nodata  \\ 
$ sidelobeThreshold=2.5 $ &   736 &   7.1 & 4.622e-03 &  0.00 & -1.568e-04 &  0.00 & 1.333e-05 &  0.00 & 4.004e-05 &   0.00\\ 
$ sidelobeThreshold=3.5 $ &   643 &   6.4 & 4.622e-03 &  0.00 & -1.568e-04 &  0.00 & 1.333e-05 &  0.00 & 4.004e-05 &   0.00\\ 
$ noiseThreshold=4.5 $ &   687 &   0.0 & 4.622e-03 &  0.00 & -1.568e-04 &  0.00 & 1.333e-05 &  0.00 & 4.004e-05 &   0.00\\ 
$ noiseThreshold=5.5 $ &   687 &   0.0 & 4.622e-03 &  0.00 & -1.568e-04 &  0.00 & 1.333e-05 &  0.00 & 4.004e-05 &   0.00\\ 
$ lowNoiseThreshold=1.0 $ &   687 &   0.0 & 4.622e-03 &  0.00 & -1.568e-04 &  0.00 & 1.333e-05 &  0.00 & 4.004e-05 &   0.00\\ 
$ lowNoiseThreshold=2.0 $ &   687 &   0.0 & 4.622e-03 &  0.00 & -1.568e-04 &  0.00 & 1.333e-05 &  0.00 & 4.004e-05 &   0.00\\ 
\cutinhead{ W0116 } 
 Control &   277 & \nodata & 8.555e-03 & \nodata & -5.160e-04 &  \nodata  & 1.499e-06 & \nodata & 1.179e-04 & \nodata  \\ 
$ sidelobeThreshold=2.5 $ &   304 &   9.7 & 8.555e-03 &  0.00 & -5.160e-04 &  0.00 & 1.499e-06 &  0.00 & 1.179e-04 &   0.00\\ 
$ sidelobeThreshold=3.5 $ &   256 &   7.6 & 8.555e-03 &  0.00 & -5.160e-04 &  0.00 & 1.499e-06 &  0.00 & 1.179e-04 &   0.00\\ 
$ noiseThreshold=4.5 $ &   277 &   0.0 & 8.555e-03 &  0.00 & -5.160e-04 &  0.00 & 1.499e-06 &  0.00 & 1.179e-04 &   0.00\\ 
$ noiseThreshold=5.5 $ &   277 &   0.0 & 8.555e-03 &  0.00 & -5.160e-04 &  0.00 & 1.499e-06 &  0.00 & 1.179e-04 &   0.00\\ 
$ lowNoiseThreshold=1.0 $ &   277 &   0.0 & 8.555e-03 &  0.00 & -5.160e-04 &  0.00 & 1.499e-06 &  0.00 & 1.179e-04 &   0.00\\ 
$ lowNoiseThreshold=2.0 $ &   277 &   0.0 & 8.555e-03 &  0.00 & -5.160e-04 &  0.00 & 1.499e-06 &  0.00 & 1.179e-04 &   0.00\\ 
\cutinhead{ NGC1068 } 
 Control &  2400 & \nodata & 8.830e-03 & \nodata & -3.638e-04 &  \nodata  & -1.456e-06 & \nodata & 7.318e-05 & \nodata  \\ 
$ sidelobeThreshold=2.5 $ & 12673 & 428.0 & 8.786e-03 &  0.50 & -3.354e-04 &  7.81 & -8.325e-07 & 42.80 & 7.065e-05 &   3.45\\ 
$ sidelobeThreshold=3.5 $ &  2353 &   2.0 & 8.825e-03 &  0.06 & -3.660e-04 &  0.60 & -1.460e-06 &  0.34 & 7.319e-05 &   0.02\\ 
$ noiseThreshold=4.5 $ &  4002 &  71.3 & 8.792e-03 &  0.43 & -3.543e-04 &  2.61 & -1.285e-06 & 11.74 & 7.226e-05 &   1.25\\ 
$ noiseThreshold=5.5 $ &  1734 &  27.8 & 8.854e-03 &  0.27 & -3.775e-04 &  3.77 & -1.618e-06 & 11.17 & 7.380e-05 &   0.85\\ 
$ lowNoiseThreshold=1.0 $ &  2400 &   0.0 & 8.830e-03 &  0.00 & -3.638e-04 &  0.00 & -1.456e-06 &  0.00 & 7.318e-05 &   0.00\\ 
$ lowNoiseThreshold=2.0 $ &  2400 &   0.0 & 8.830e-03 &  0.00 & -3.638e-04 &  0.00 & -1.456e-06 &  0.00 & 7.318e-05 &   0.00\\ 
\cutinhead{ NGC6334I } 
 Control &  3815 & \nodata & 4.692e+00 & \nodata & -7.766e-02 &  \nodata  & 2.401e-02 & \nodata & 2.711e-02 & \nodata  \\ 
$ sidelobeThreshold=2.5 $ &  4027 &   5.6 & 4.691e+00 &  0.04 & -7.743e-02 &  0.29 & 2.398e-02 &  0.12 & 2.712e-02 &   0.04\\ 
$ sidelobeThreshold=3.5 $ &  3223 &  15.5 & 4.690e+00 &  0.06 & -7.914e-02 &  1.92 & 2.384e-02 &  0.72 & 2.712e-02 &   0.04\\ 
$ noiseThreshold=4.5 $ &  3815 &   0.0 & 4.692e+00 &  0.00 & -7.766e-02 &  0.00 & 2.401e-02 &  0.00 & 2.711e-02 &   0.00\\ 
$ noiseThreshold=5.5 $ &  3815 &   0.0 & 4.692e+00 &  0.00 & -7.766e-02 &  0.00 & 2.401e-02 &  0.00 & 2.711e-02 &   0.00\\ 
$ lowNoiseThreshold=1.0 $ &  3815 &   0.0 & 4.692e+00 &  0.00 & -7.766e-02 &  0.00 & 2.401e-02 &  0.00 & 2.711e-02 &   0.00\\ 
$ lowNoiseThreshold=2.0 $ &  3248 &  14.9 & 4.690e+00 &  0.05 & -7.934e-02 &  2.16 & 2.381e-02 &  0.83 & 2.716e-02 &   0.20\\ 
\enddata
\tablecomments{Column (1): Image. Column (2): Number of pixels in the mask. Column (3): Percent difference in the absolute number of pixels in mask for the test and control images. Column (4): Maximum value in final image. Column (5): Percent difference between maximum value in test and control images. Column (6): Minimum value in final image. Column (7): Percent difference between the minimum value in the test and control images. Column (8): Mean value in the final image. Column (9): Percent difference between the mean value in the test and control images. Column (10): Noise in final image (estimated using the median absolute deviation). Column (11): Percent difference between the noise in the test and control images. }
\end{deluxetable}

The W0116 and SPT0346 data sets provide an excellent demonstration of the influence of the {\it sidelobeThreshold} parameter. These images both include only a single point source. The point source emission is masked initially using either the {\it sidelobeThreshold} or {\it noiseThreshold} (depending on the major cycle). After the first major cycle, however, the masks are cascaded down to lower signal-to-noise with the threshold being set by the {\it sidelobeThresholdValue} (not the {\it lowNoiseThresholdValue}). Therefore, the entire mask in both cases is determined by the {\it sidelobeThreshold} parameter. The {\it noiseThresold} parameter only plays a role in the intermediate major cycles and the {\it lowNoiseThreshold} value is not used.

\begin{figure}
\centering
\includegraphics[width=\textwidth]{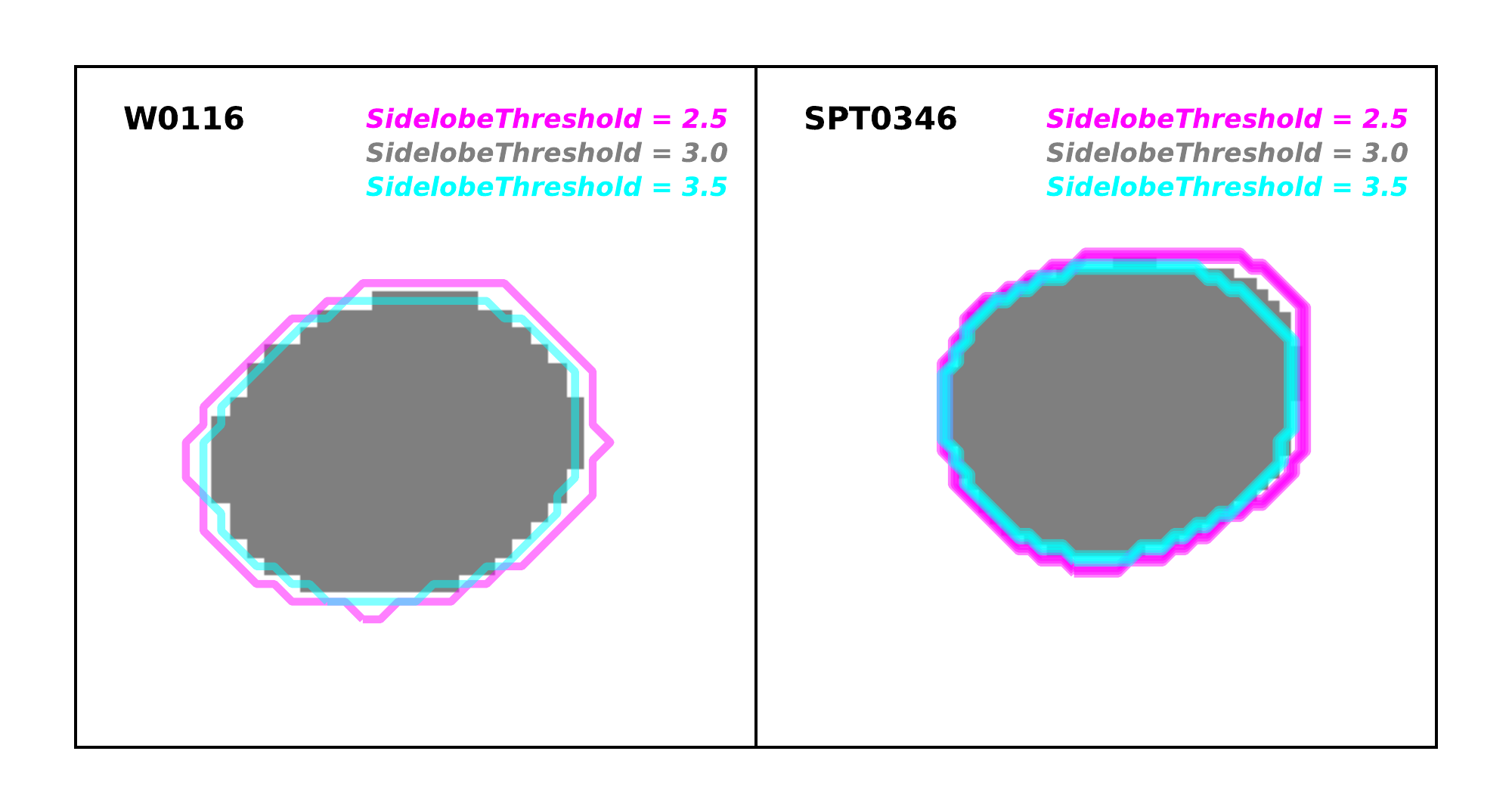}
\caption{Masks generated for W0116 (left) and SPT0346 (right) data sets for three different values of the {\it sidelobeThreshold}. The gray region shows the mask generated by our nominal parameter value. The magenta and cyan contours show the masks generated by decreasing the {\it sidelobeThreshold} to 2.5 and increasing the {\it sidelobeThreshold} to 3.5, respectively. The changes in the masks generated by varying the {\it sidelobeThreshold} are minor for these cases.} \label{fig:parameter_variation_W0116_SPT0346}
\end{figure}

The NGC6334I data set has similar behavior to the W0116 and SPT0346 data sets. The initial regions are picked out via the {\it sidelobeThreshold} and {\it noiseThreshold} parameters. Then the mask is cascaded down to low signal-to-noise using the {\it sidelobeThreshold}, except for the case where the {\it lowNoiseThreshold }is set equal to 2.0. In that case, the low noise threshold mask is set by the {\it lowNoiseThreshold}, not the {\it sidelobeThreshold}. We also note that the $sidelobeThreshold = 3.5$ and the $lowNoiseThreshold = 2.0$ masks are very similar, although not identical. This result demonstrates that masks generated by \textsc{auto-multithresh} are not unique to a particular set of parameters, but that similar masks can be generated with different sets of parameters. 

\begin{figure}
\centering
\includegraphics[width=\textwidth]{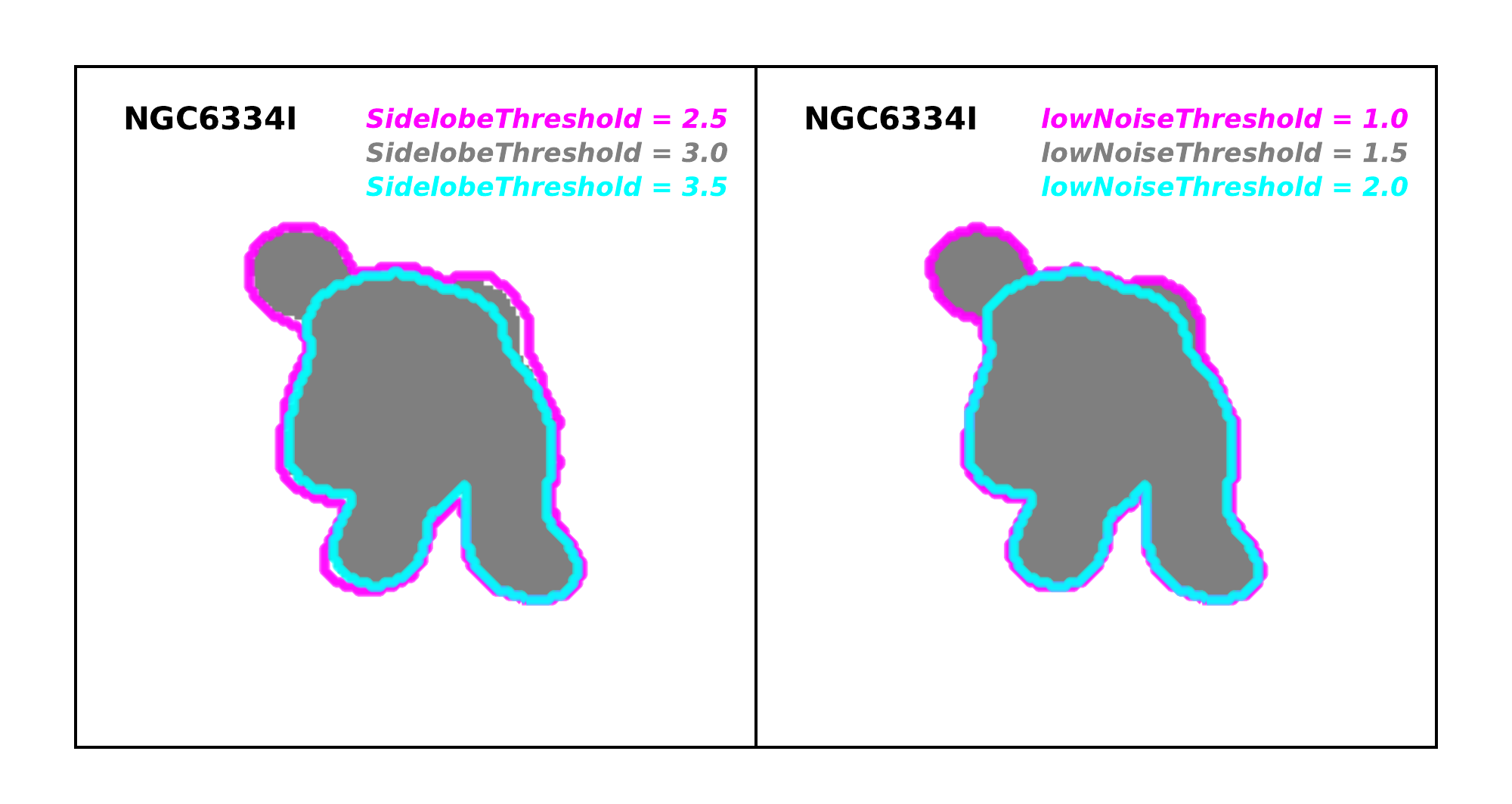}
\caption{Masks generated for the NGC6334I test data set for different parameter values. The left figure shows the masks generated by our nominal {\it sidelobeThreshold} value of 3.0 (gray region), decreasing the {\it sidelobeThreshold} to 2.5 (magenta contour), and increasing the {\it sidelobeThreshold} to 3.5 (cyan contour). All three masks are very similar. Decreasing the {\it sidelobeThreshold} results in a slightly more extended mask with one additional region included. The right figure shows the masks generated by our nominal {\it lowNoiseThreshold} value of 1.5 (gray region), decreasing the {\it lowNoiseThreshold} to 1.0 (magenta contours), increasing the {\it lowNoiseThreshold} to 2.0 (cyan contours). The masks for {\it lowNoiseThreshold} equal to 1.0 and 1.5 are identical because the {\it sidelobeThreshold} determines the low signal-to-noise mask. Increasing the {\it lowNoiseThreshold} to 2.0 causes this parameter to determine the low signal-to-noise mask resulting in a less extended mask.} \label{fig:parameter_variation_NGC6334}
\end{figure}

\begin{figure}
\centering
\includegraphics[width=\textwidth]{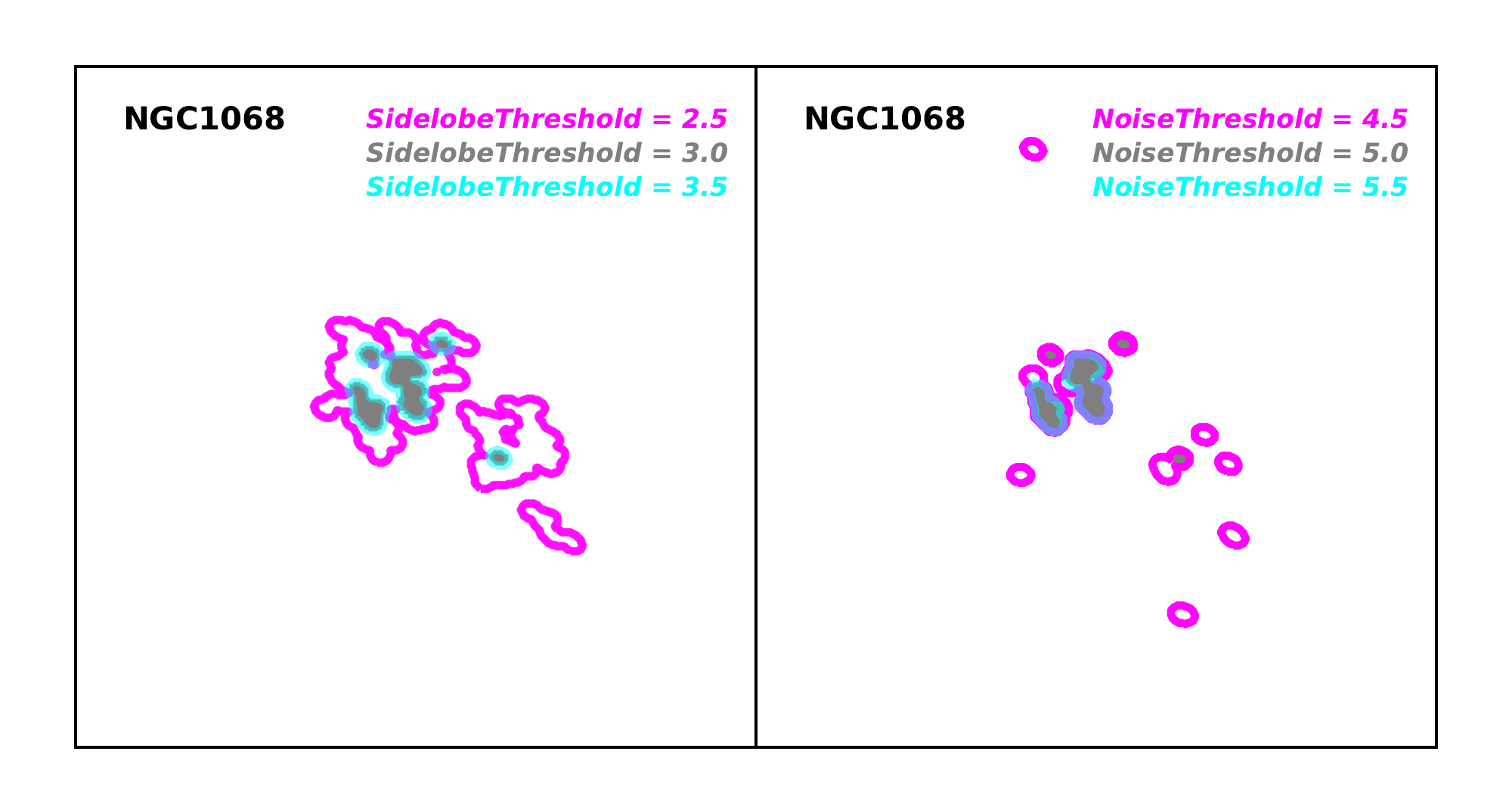}
\caption{Masks for the NGC1068 test data set for different parameter values. The left image shows the effects of changing the {\it sidelobeThreshold} for this data set. The nominal {\it sidelobeThreshold} of 3.0 is shown as a gray region with a {\it sidelobeThreshold} of 2.5 shown as magenta contour and a {\it sidelobeThrehsold} of 3.5 as a cyan contour. Decreasing the {\it sidelobeThreshold} results in significantly larger mask since it controls the cascade down to lower signal-to-noise. The right figure shows the effects of changing the {\it noiseThreshold} for this test case. Again, our nominal {\it noiseThreshold} value of 5.0 is shown as gray regions, the decreased {\it noiseThreshold} of 4.5 as a magenta contour, and the increased {\it noiseThreshold} of 5.5 as a cyan contour. Decreasing the {\it noiseThreshold} from nominal in this case results in more spurious regions in the image. However, the changes in the mask are smaller than the changes generated by changing the {\it sidelobeThreshold}.} \label{fig:parameter_variation_NGC1068}
\end{figure}

Changing the {\it noiseThreshold} only affects the masks produced for the NGC1068 image. This image is dominated by artifacts due to issues with the underlying visibility data, not sidelobes.  Decreasing the {\it noiseThreshold} picks up more artifacts. This parameter will likely play a more important role in images with a large number of point-like sources (e.g., 1.4GHz radio continuum images with high (arcsec) resolution), where fainter and fainter sources are revealed as the image is cleaned. As Figure~\ref{fig:parameter_variation_NGC1068} demonstrates, however, the changes in the mask generated by changing the {\it noiseThreshold} are less than the changes in the mask generated by changing the {\it sidelobeThreshold}.

Based on the image statistics in Table~\ref{tab:parameter_variation}, the changes to the mask parameters appear to have the most effect on the final images for the data sets with complex emission (NGC6334I) or significant artifacts (NGC1068) with the most significant differences for the latter case. The simple imaging cases (W0116 and SPT0346) show no difference in the resulting images for different mask parameters. While a fraction of the ALMA images are as simple as W0116 and SPT0346, many of the data sets produced by ALMA have complex extended emission like those seen in NGC6334I. In addition, while the data quality from ALMA is very high, there are inescapable limitations based on weather, equipment, etc. that means that the data is not perfect. Thus, in operational situations where a range of emission morphologies and data quality may be encountered it is important to tune the \textsc{auto-multithresh} parameter to produce the best images. Finally, we note that the overall quality of the mask is much more important for self-calibration, which requires a very accurate model of the sky brightness to avoid introducing artifacts into the final image. Thus, the minor differences in the resulting images become much more important, especially for faint, extended emission. The \textsc{auto-multithresh} masking processes are more stringent than one might needs for a simple imaging-only case, so that in the future the ALMA Imaging Pipeline can support automated self-calibration.


\subsection{Speed} \label{sec:speed}

We imaged a suite of ALMA benchmark data sets using the standard ALMA Imaging Pipeline procedures and parameters to show how long each step in the algorithm takes for typical ALMA data sets. To do this, we used the North American ALMA Science Center (NAASC) cluster. At the time of these tests, the cluster consisted of two models of servers:
Dell PowerEdge R430 Intel(R) Xeon(R) CPU E5-2640 v3 @ 2.60GHz
and Dell PowerEdge R620 Intel(R) Xeon(R) CPU E5-2670 0 @ 2.60GHz.
The servers all had 16 available cores and 252 GB available memory and were running Red Hat 6.9. The servers were connected to a Lustre system consisting of  1 Supermicro MDS node and 9 Supermicro OSS nodes. The Lustre server version was 2.5.5,  ran Red Hat 6.8 and used the ldiskfs/ext4 backend file system. Each data set was imaged in serial mode on a single node. 

\begin{figure}
    \centering
    \includegraphics[width=0.5\textwidth]{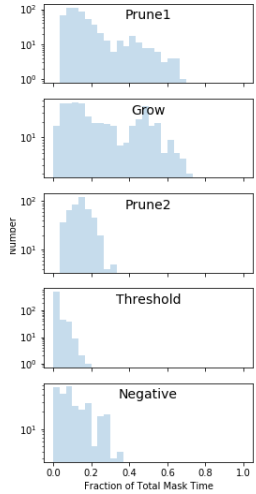}
    \caption{Fraction of total mask time spent in each step in the \textsc{auto-multithresh} algorithm for each major cycle for all data sets in the ALMA benchmark. The prune and grow steps in the algorithm are the most time-consuming. }
    \label{fig:timing_results}
\end{figure}

Figure~\ref{fig:timing_results} shows how much of the total masking time was spent on each step of the algorithm for each major cycle for the ALMA benchmark data sets. For cubes, this represents the total masking time for all channels in one major cycle. We have removed images with very short masking times ($< 10$s) to avoid obtaining unrealistically large fractions. Both the initial threshold and the negative threshold steps take approximately a quarter of the masking time. The initial threshold step includes the noise calculation, which can take up to two times longer for the improved noise calculation described in Appendix~\ref{sec:noise} over the original simple median absolute deviation estimate used for these tests. The most time consuming steps are the two pruning steps and the grow step. These two steps involve iterating over all pixels in an image, which can be very expensive for especially for large images. Although the initial prune and the mask growth steps are essential for creating a good mask, the pruning of the low signal-to-noise threshold mask does not generally have a significant effect on the mask.  We have included a parameter ({\it doGrowPrune}) in the {\tt tclean} \textsc{auto-multithresh} implemention to turn this feature off to increase the speed. The {\it minPercentChange} parameter also significantly improves the overall time it takes to clean an image using \textsc{auto-multithresh} by stopping future evaluation of new masks when the mask stops changing significantly. 

\section{Comparison with Existing Auto-masking Algorithms} \label{sec:comparison}

A number of other automated masking algorithms have been developed previously, typically for specific projects. These algorithms include one developed for OBIT \citep{Cotton2007EVLAWindowing}, one developed in AIPS \citep{Greisen2009AIPSAIPS} and several previous versions developed for CASA including the task {\tt boxit} \citep{Kimball2011AutomatedCASA}, \textsc{auto-thresh2} (and its close relative \textsc{auto-thresh}) in the {\tt tclean} task \footnote{\url{https://casa.nrao.edu/casadocs/casa-5-1.2/synthesis-imaging/masks-for-deconvolution}}, and a heuristic developed for the M100 ALMA CASA guide\footnote{\url{https://casaguides.nrao.edu/index.php/M100_Band3}}. 

Although the other automasking algorithms differ considerably in detail, they rely on similar principles to the \textsc{auto-multithresh} algorithm described in Section~\ref{sec:description}. In general, the algorithms attempt to select out the most significant regions of emission either by regions above a certain RMS threshold (OBIT automasking, AIPS automasking, {\tt boxit}, M100 CASA guide, \textsc{auto-thresh2}), or regions above a certain fraction of the peak residual (\textsc{auto-thresh2}, AIPS automasking and {\tt boxit}). Both  AIPS automasking and {\tt boxit} select islands of low signal-to-noise emission surrounding significant emission, which is similar in effect to the grow operation that \textsc{auto-multithresh} does. The pruning operation for the \textsc{auto-multithresh} prototype was taken directly from the M100 CASA guide. The \textsc{auto-multithresh} algorithm also adopted some ideas from early versions of automasking in the VLA Sky Survey Imaging Pipeline including taking the maximum of the noise and sidelobe thresholds and smoothing and cutting the mask to enlarge it (S. Myers, private communication).


The \textsc{auto-multithresh} algorithm has several advantages over these earlier algorithms. First, the parameters controlling these algorithms are only partially specified  relative to fundamental properties of the image. The \textsc{auto-multithresh} algorithm parameters are entirely general, which means that they usually do not need to be changed significantly for each data set being imaged. For example, the {\it noiseThreshold} and {\it lowNoiseThreshold} are specified in terms of signal-to-noise rather than in units of Jy~beam$^{-1}$,  the {\it sidelobeThreshold} is specified relative to the sidelobe level in the image, and the {\it minBeamFrac} parameter is specified relative to the beam area. Second, it uses pixel masks rather than regions which improves the masking of complex images, in particular those typically encountered by ALMA. The OBIT and  AIPS autoboxing routines and {\tt boxit} task produce a set of mask regions with fixed positions, shapes, and sizes. It can be difficult to construct a mask for complex emission with a set of regions, but a pixel mask is able to capture complex emission easily. Finally, the \textsc{auto-multithresh} has been systematically validated on a wide variety of images rather than a few test cases to ensure that it can handle a wide variety of image morphologies. It has been run on all data that have been processed by the ALMA Pipeline starting in Cycle 5. This represents 70\% of all ALMA projects in Cycle 5 and 94\% of all ALMA projects in Cycle 6.  To the authors' knowledge, none of the previous auto-masking algorithms have been used on this scale.

The downside to this generality and flexibility is that the \textsc{auto-multithresh} algorithm is complex. We have five main parameters that have to be tuned for different use cases and the masking process itself is fairly time-consuming. Once one gains experience with the algorithm, adjusting these parameters is relatively straightforward, but experience has shown that new users generally need assistance adjusting the parameters. We have included advice on how to tune the \textsc{auto-multithresh} parameters in Appendix~\ref{sec:tuning}.

All the automated masking algorithms described above, including \textsc{auto-multithresh} work in parallel to the minor cycle. In the future, it may be preferable to build any automated masking or signal identification algorithms directly into the minor cycle itself since they have similar goals (identifying regions of significant emission). An example of this approach is the so-called non-amnesiac clean  \citep{Golap2015Non-AmnesicAlgorithm}. In this probabilistic approach, clean components are preferentially placed near other previously determined clean components during the deconvolution process.

The algorithms described above all use a top-down approach for identifying emission, which produces a more restrictive mask. While this approach works well when modeling sources as a collection of point sources, it may not be ideal with deconvolving images using multiple scales (i.e., the so-called multi-scale clean) as the mask produced is often too restrictive for the largest scale. In that case, an outside-in approach using a large, relatively unrestricted mask is more appropriate. For this reason, the PHANGS-ALMA survey (Leroy et al, in prep) has adopted an alternate approach, which uses a heavily tapered initial image to create a mask for a subsequent multi-scale clean that cleans down to 4 to 5 $\sigma$ and then finishes with a more restrictive mask for a single scale clean. Images produced using this method can be found in \citet{Sun2018Cloud-scaleGalaxies}.

\section{Summary and Conclusions} \label{sec:summary}

We have developed an automated masking algorithm that operates every major cycle within \textsc{clean} to identify regions of significant emission. This algorithm, which we refer to as \textsc{auto-multithresh},  masks significant peaks based on the sidelobe and noise levels in the images and then cascades these masks down to lower signal-to-noise levels.   The algorithm was implemented in CASA and has been used in production by the ALMA Imaging Pipeline since Cycle 5. Based on the results of production use of the algorithm, we have met our initial goal of improving the quality of images produced by the ALMA Imaging Pipeline. However, the algorithm is sufficiently general that it has been successfully used with data from other interferometers like the VLA and ATCA.

The \textsc{auto-multithresh} input parameters are specified with respect to the fundamental properties of the image, allowing us to use the same set of parameters for many images. They are moderately sensitive to the point spread function (PSF) of the data: data with poor PSFs (i.e., poor {\it u-v} coverage) need slightly different parameters than data with good PSFs (i.e., good {\it u-v} coverage). Three different sets of parameters are able to image the wide variety of ALMA data.

Because the algorithm operates every major cycle, it can significantly increase the amount of time spent cleaning an image. The slowest portions of the algorithm are the pruning of the small (likely spurious) regions and the binary dilation of the initial mask down to low signal-to-noise. In the CASA implementation, we have introduced parameters that reduce how often these steps occur.

The \textsc{auto-multithresh} algorithm shares some common features with other automated masked algorithms that have been developed previously, but is more general. This generality allows it to function successfully as part of a pipeline, but introduces additional complexity into the algorithm. Embedding the algorithm more deeply into the minor cycle could provide further performance improvements.

\acknowledgments

The authors would like to thank the referee, Erik Rosolowsky, for his helpful comments. AAK would like to thank Urvashi Rau for showing her how to modify {\tt tclean} to include masking, Steve Myers for sharing some early ideas for auto-masking in the VLASS Pipeline, CJ Allen for unearthing historic NAASC cluster information, and NAASC software support team in its various incarnations.  The National Radio Astronomy Observatory is a facility of the National Science Foundation operated under cooperative agreement by Associated Universities, Inc. This paper makes use of the following ALMA data:  ADS/JAO.ALMA\#2013.1.00014.S, ADS/JAO.ALMA\#2013.1.00576.S, ADS/JAO.ALMA\#2013.1.00722.S, and ADS/JAO.ALMA\#2015.1.00150.S, ADS/JAO.ALMA\#2017.1.00964.S, and ADS/JAO.ALMA\#2016.1.00314.S. ALMA is a partnership of ESO (representing its member states), NSF (USA) and NINS (Japan), together with NRC (Canada), MOST and ASIAA (Taiwan), and KASI (Republic of Korea), in cooperation with the Republic of Chile. The Joint ALMA Observatory is operated by ESO, AUI/NRAO and NAOJ. 

\appendix

\section{Noise Estimation} \label{sec:noise}

There can be significant signal in residual images, especially during the first few major cycles of clean. The presence of this signal requires that we use noise estimators that are relatively unaffected by outliers, commonly referred to as  ``robust'' estimators. In  implementations of \textsc{auto-multithresh} prior to CASA 5.5, we estimated the noise using the median absolute deviation or $MAD$, which is defined as 
\begin{equation}
MAD = median | x - median(x) |
\end{equation}
where $x$ is an array of data values \citep{Iglewicz1983RobustLocation,Rousseeuw1993AlternativesDeviation,Beers1990MeasuresApproach}. We multiplied the $MAD$ by a factor of 1.4826 to make this statistic equivalent to the width of the Gaussian distribution (i.e., the Gaussian $\sigma$). This noise estimate is used when the parameter {\it fastNoise} is set equal to True.

\begin{figure}
    \centering
    \includegraphics[width=\textwidth]{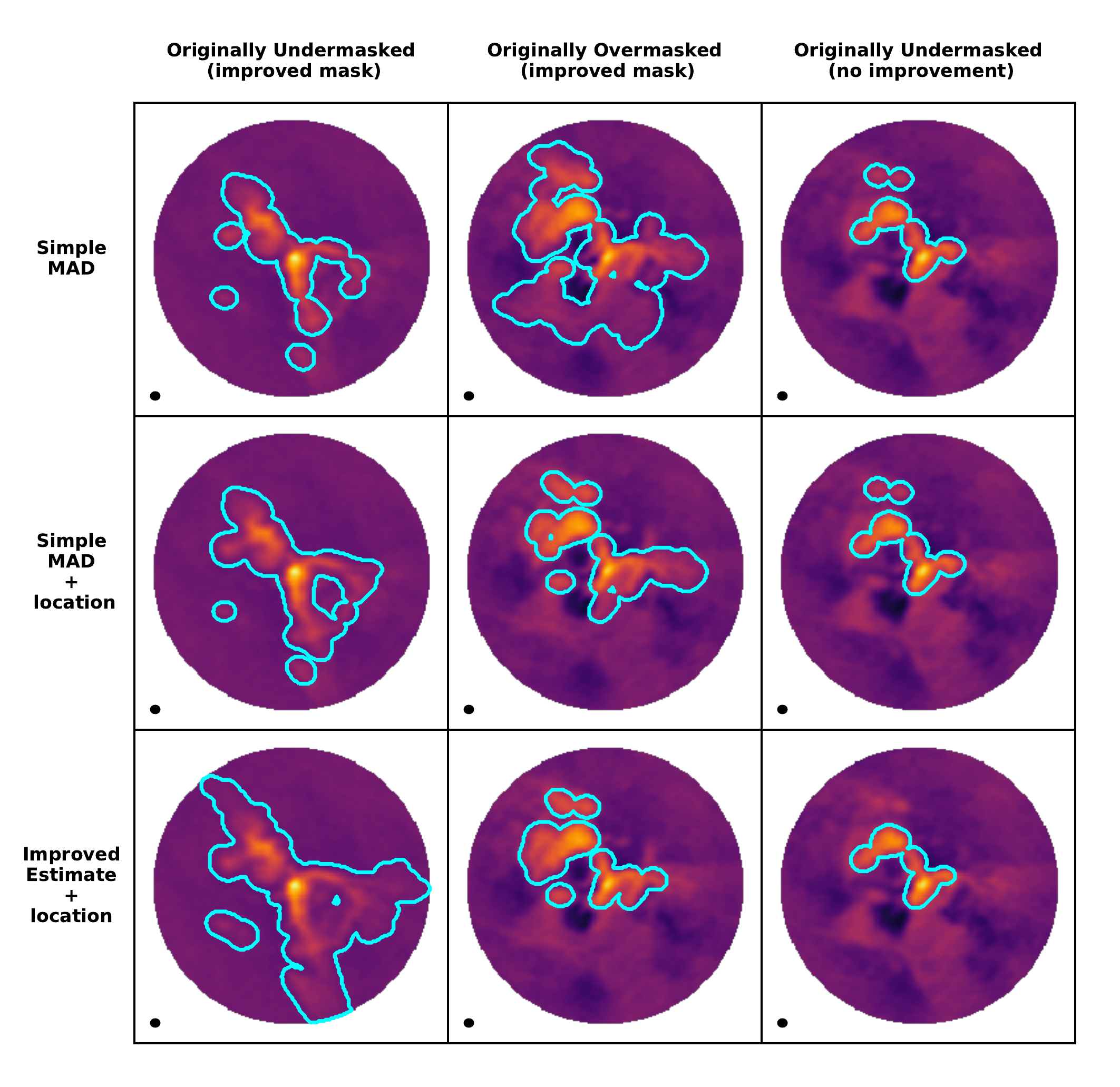}
    \caption{A comparison of the masks generated using different noise estimates for three channels in a particularly problematic cube. The top row shows the mask created using the original noise estimate (median absolute deviation or MAD). The left and right panels show two channels that were undermasked and the middle column shows a channel that was overmasked. All the images have the same range and stretch. The middle row shows the mask created by using the MAD and including the central location of the data (estimated using the median) in the threshold calculation for the same three channels as above. Including the location improves both the undermasked and overmasked cases in the left and middle panel and leaves the undermasked case in the right panel unchanged. Finally, the bottom row shows the masks generated using the improved noise estimated described in Appendix~\ref{sec:noise} and includeing the central location of the data in the threshold calculation for the same three channels as above. The masking in the left and middle panels was improved by a combination of the new noise estimate and including a central location for the data. The masking in the right panel was not made significantly worse by this combination.}
    \label{fig:noise_example}
\end{figure}

The $MAD$ is a popular robust estimator of the noise because it is relatively simple to compute and has a high breakdown point (0.5). The latter feature means that up to half the values in a data set can be outliers and the $MAD$ will still provide a good estimate of the noise in the image. This assumption holds for many, but not all, of the datasets we have tested. ALMA's relatively narrow field of view combined with the extended molecular emission seen in many Galactic star forming regions leads to a significant number of residual images where most of the field of view of the initial residual image has signal. In this case, the $MAD$ overestimates the noise in the residual image and thus does not mask all the significant emission in the resulting residual image. The top right panel of Figure~\ref{fig:noise_example} shows an example of this behavior.


Our initial assumption that the distribution of pixel values is symmetric about zero can also lead to poor masking. In cases with significant absorption or emission, the distribution can be centered around a non-zero value, which we refer to as the ``location'' of the data following standard statistics terminology. Appropriately identifying outliers from the distribution (i.e., signal) in these cases requires taking into account the location of the distribution. Ignoring this effect results leads to undermasking of positive features and overmasking of negative features for locations greater than zero, and the opposite effects for locations less than zero. The top left and top middle panels of Figure~\ref{fig:noise_example} show an example of undermasked and overmasked images, respectively. 

The \textsc{auto-multithresh} implementations in CASA 5.5 and higher include an estimate of the location in the threshold calculations and optionally provide a more robust estimate of the noise. The new noise estimate is used when the parameter {\it fastNoise} is set equal to False. It calculates the noise in two different ways depending on whether or not a mask is present. If there is no mask present, then outliers are removed from the pixel distribution using Chauvenet's criterion (\citealp{Peirce1852CriterionObservations}, \citeauthor{Chauvenet1891AII} 1852/1960).\footnote{Although this criterion is typically referred to as Chauvenet's criterion, the original source is \citet{Peirce1852CriterionObservations}. \citeauthor{Chauvenet1891AII} 1852/1960  directly quotes \citet{Peirce1852CriterionObservations}.}. This outlier rejection technique assumes that the data are normally distributed and iteratively removes values that are statistically unlikely for a normal distribution. For more information on the CASA implementation see \url{https://casa.nrao.edu/casadocs/casa-5.4.0/global-task-list/task_imstat/about}. If there is a mask present, the noise is estimated via the re-scaled MAD using the pixels of the region outside the  clean mask, but inside the primary beam mask. We refer to this estimate as the masked MAD. This region excludes the regions of significant emission that have already been found by \textsc{auto-multithresh} but that have potentially not been adequately cleaned. In both cases, we estimate the location of the data using the median, which is a robust estimator for this quantity, and add it to the threshold. The data used to calculate the median is the same data set used to estimate the noise, that is, either the resulting data set after  outliers have been removed using Chauvenet's criterion or the data outside the clean mask, but inside the primary beam mask.

Typically more sophisticated modern robust scale estimators like the biweight \citep{Hoaglin1983UnderstandingAnalysis,Beers1990MeasuresApproach} are preferred over Chauvenet's criterion, which was originally developed in the mid-1800s. However, our testing has shown that because the Chauvenet's criterion assumes a normal distribution it provides a better estimate of the noise for our use case than robust estimators like the biweight which only assume that the distribution is symmetric.

Figure~\ref{fig:noise_example} shows the improvements in the masks for one of the more problematic data sets (a Galactic star-forming region with a widespread outflow) generated by including an estimate of the location in the threshold and by using a more sophisticated noise estimate. For two out of the three cases, the mask generated via the latter procedure is a significant improvement on the original mask (bottom left and bottom middle panels). However, it is not a panacea since we still have channels are that undermasked (bottom right panel). 

Fundamentally we are limited here by our ability to estimate the noise in a signal-rich image and by the significant negative bowls generated by resolved out emission in this image. We could potentially use other signal-free channels to estimate the noise, but this requires making the assumption that the noise for each channel is roughly similar, which may not be the case for many ALMA images due to the presence of atmospheric lines. The inclusion of data from a more compact array or single dish observations would likely  improve the masking because it would reduce or eliminate the negative bowls widening and skewing the distribution of pixel values. For cases like those in bottom right panel, the masks produced by \textsc{auto-multithresh} can be combined with masks produced manually either interactively or by running {\tt tclean} in restart mode twice, once with a user mask and once with \textsc{auto-multithresh}.

\section{Pruning Implementation} \label{sec:prune_implementation}

This appendix presents technical details on the pruning algorithm used by \textsc{auto-multithresh}. This step is one of the two most time-consuming steps in the \textsc{auto-multithresh} algorithm (the other being the binary dilation of the mask). Pruning operates on an intermediate mask produced by the \textsc{auto-multithresh} algorithm, which is stored as a pixel mask image. To prune the image, each masked region is identified by scanning the mask image using the Depth-First Search algorithm, which is commonly used in connected component labelling. For each pixel, all adjacent neighboring pixels are checked if they are masked and continued to their adjacent pixels until it reaches to an unmasked pixel. The visited pixels are marked as visited. Then the algorithm  continues on to the next "unvisited" pixel. In this scheme, a matrix of the size of mask image keeps track of label indices for unique regions. The size of each region is extracted from the label matrix since it indexed with unique integer numbers. In the second step, pruning of regions is done by using the size information and the label matrix as reference for region locations since it maps pixel by pixel to the mask image.

\section{Tuning the \textsc{auto-multithresh} Parameters} \label{sec:tuning}

As discussed in Section~\ref{sec:input_params}, the \textsc{auto-multithresh} parameters should adjusted based on the quality of the PSF: PSFs with higher sidelobes will need a different set of parameters than PSFs with lower or minimal sidelobes. Tuning the \textsc{auto-multithresh} parameters generally requires some experimentation. In this Appendix, we present guidelines for how to most efficiently adjust these parameters for use cases beyond the ALMA Pipeline parameters presented in Table~\ref{tab:pipeline_params}.

The four key \textsc{auto-multithesh} parameters to adjust are the {\it noiseThreshold}, {\it sidelobethreshold}, {\it lowNoiseThreshold}, and the {\it minBeamFrac}. Practically speaking, adjusting the {\it cutThreshold} and {\it smoothFactor} is not necessary.  We suggest selecting a small test case and first adjusting {\it noiseThreshold} and {\it sidelobeThreshold} to capture the emission peaks. In this step, you may need to adjust {\it minBeamFrac} to optimize which regions are retained. After adjusting {\it noiseThreshold}, {\it sidelobeThreshold}, and {\it minBeamFrac} parameters, then adjust {\it lowNoisethreshold} to capture the extended emission. In some cases, you may need to adjust the {\it sidelobeThreshold} to mask the extended emission since the threshold for the extended emission is set by the maximum of the {\it lowNoiseThresholdValue} and {\it sidelobeThresholdValue}  (see Section~\ref{tab:parameter_variation} for an example). 

There are several ways one can gauge how the masking is proceeding. First, if you are running in serial mode, you can turn on interactive cleaning mode in {\tt tclean} ($interactive=True$). In this mode, the mask will be displayed in the viewer after every major cycle.  You can also modify the mask as it is being produced in this mode, which is useful in cases where the noise is difficult to estimate. Note that the number of cycle iterations must be set equal to the number of iterations to produce the same mask in both interactive and non-interactive mode. Second, you can turn on verbose mode for \textsc{auto-multithresh} ($verbose=True$). This mode writes information to the logger about what threshold is being used in what channel for every major cycle as well as information on the {\it noiseThresholdValue} and {\it sidelobeThresholdValue} and the number of regions found and pruned. Finally, you can inspect the final mask in the viewer. The most useful configuration is two panels -- one with the image and one with the residual -- with the mask overlaid as contours.

For more complex masking cases, you can also build up masks via multiple calls to {\tt tclean}. Two example cases where this might be useful are for a) faint continuum with spectral line emission and b) high dynamic range images with many point sources. For the first case, one approach would be to run {\tt tclean} once creating a user mask where the continuum emission is and then run it a second time, taking the user mask as input but turning on \textsc{auto-multithresh} to further expand the mask. For the second case, a suggested approach would be to run {\tt tclean} once with a small number of iterations and pruning turned off ({\it minBeamFrac}=0.0) to mask the brightest sources. After the initial mask, restart {\tt tclean} with {\it minBeamFrac} set to the typical value to avoid masking noise peaks and clean to the threshold.







\vspace{5mm}


\software{astropy (various), CASA ($\geq$4.7.1), matplotlib, aplpy}

\facility{ALMA}


\end{document}